%% file: Main.tex
\documentclass[a4paper]{IEEEtran}

\usepackage[margin=1.4cm]{geometry}

\input{Packages}

\input{Commands}

\input{Acronyms}

\begin{document}


\title{On the Properties of the Compound Nodal Admittance Matrix of Polyphase Power Systems}

\author{
	Andreas~Martin~Kettner,~\IEEEmembership{Member,~IEEE},~and~Mario~Paolone,~\IEEEmembership{Senior~Member,~IEEE}%
	\thanks{This work was supported by the Swiss National Science Foundation through the National Research Programme NRP-70 ``Energy Turnaround''.}%
	\thanks{The authors are with the Swiss Federal Institute of Technology of Lausanne (EPFL), CH-1015 Lausanne, Switzerland. E-mail: forename.surname@epfl.ch.}%
}	




\maketitle

\input{"Sections/Abstract"}




\input{Sections/Introduction}

\input{Sections/Foundations}

\input{Sections/Properties}

\input{Sections/Implications}

\input{Sections/Conclusions}

\appendices

\input{Sections/Equipment}




\bibliographystyle{IEEEtran}
\bibliography{Bibliography}



\input{Sections/Biographies}

\end{document}

%% file: Packages.tex
\usepackage{cite}

\ifCLASSINFOpdf
	\usepackage[pdftex]{graphicx}	
\else
	\usepackage[dvips]{graphicx}
\fi
\usepackage[cmex10]{amsmath}
\usepackage{amssymb}
\usepackage{amsthm}
\usepackage{mathtools}	
\usepackage{mathrsfs}	
\usepackage[cal=euler,scr=rsfso]{mathalfa}	

\usepackage{array}

\ifCLASSOPTIONcompsoc
	\usepackage[caption=false,font=normalsize,labelfont=sf,textfont=sf]{subfig}
\else
	\usepackage[caption=false,font=footnotesize]{subfig}
\fi

\usepackage{url}

\usepackage{subdepth}	
\usepackage{xspace}	

\usepackage{tikz}
\usetikzlibrary{positioning,chains,fit,shapes,calc,decorations.pathreplacing}
\usepackage{circuitikz}

%% file: Commands.tex
\newcommand{\Tensor}[1]{\mathbf{#1}}
\newcommand{\Rank}[1]{\operatorname{rank}#1}
\newcommand{\Diagonal}[1]{\operatorname{diag}#1}
\newcommand{\Row}{\operatorname{row}}
\newcommand{\Column}{\operatorname{col}}
\newcommand{\Transpose}[1]{#1^{T}}
\newcommand{\Inverse}[1]{#1^{-1}}
\newcommand{\Schur}[2]{#1/#2}
\newcommand{\Kronecker}[2]{#1\otimes#2}

\newcommand{\Field}[1]{\mathbb{#1}}
\newcommand{\pluseq}{\mathrel{+}=}

\newcommand{\Set}[1]{\mathcal{#1}}
\newcommand{\Cardinality}[1]{|#1|}
\newcommand{\Cartesian}[2]{#1\times#2}
\newcommand{\Complement}[1]{#1_{\complement}}	

\newcommand{\Graph}[1]{\mathscr{#1}}
\newcommand{\Vertices}[1]{\operatorname{V}\left(#1\right)}
\newcommand{\Edges}[1]{\operatorname{E}\left(#1\right)}
\newcommand{\Incoming}[2]{\operatorname{E}_{\text{in}}\left(#1,#2\right)}
\newcommand{\Outgoing}[2]{\operatorname{E}_{\text{out}}\left(#1,#2\right)}
\newcommand{\Cut}[2]{\operatorname{E}_{\text{cut}}\left(#1,#2\right)}
\newcommand{\Internal}[2]{\operatorname{E}_{\text{int}}\left(#1,#2\right)}

\newcommand{\Conjugate}[1]{#1^{*}}
\newcommand{\Absolute}[1]{\left|#1\right|}
\newcommand{\Argument}[1]{\operatorname{arg}\left(#1\right)}

\newcommand{\Pie}{\Pi}				
\newcommand{\Tea}{\mathrm{T}}	

{
\theoremstyle{plain}
\newtheorem{Hypothesis}{Hypothesis}
\newtheorem{Lemma}{Lemma}
\newtheorem{Theorem}{Theorem}
\newtheorem{Corollary}{Corollary}
\newtheorem{Observation}{Observation}
}

{
\theoremstyle{definition}
\newtheorem*{Proof}{Proof}
}

\newcommand{\WhiteSpaceMagic}[1]{\vspace{#1}}

%% file: Acronyms.tex
\newcommand{\PFS}{PFS\xspace}	
\newcommand{\SE}{SE\xspace}		
\newcommand{\VSA}{VSA\xspace}	
\newcommand{\FACTS}{FACTS\xspace}	

%% file: Sections/Abstract.tex
\begin{abstract}
	Most techniques for power system analysis model the grid by exact electrical circuits.
	For instance, in power flow study, state estimation, and voltage stability assessment, the use of admittance parameters (i.e., the nodal admittance matrix) and hybrid parameters is common.
	Moreover, network reduction techniques (e.g., Kron reduction) are often applied to decrease the size of large grid models (i.e., with hundreds or thousands of state variables), thereby alleviating the computational burden.
	However, researchers normally disregard the fact that the applicability of these methods is not generally guaranteed.
	In reality, the nodal admittance must satisfy certain properties in order for hybrid parameters to exist and Kron reduction to be feasible.
	Recently, this problem was solved for the particular cases of monophase and balanced triphase grids.
	This paper investigates the general case of unbalanced polyphase grids.
	Firstly, conditions determining the rank of the so-called compound nodal admittance matrix and its diagonal subblocks are deduced from the characteristics of the electrical components and the network graph.
	Secondly, the implications of these findings concerning the feasibility of Kron reduction and the existence of hybrid parameters are discussed.
	In this regard, this paper provides a rigorous theoretical foundation for various applications in power system analysis.
\end{abstract}



%% file: Sections/Introduction.tex
\WhiteSpaceMagic{-0.2cm}

\section{Introduction}
\label{Sec:Introduction}


\IEEEPARstart{I}{nherently}, techniques for power system analysis need an exact analytical description of the grid.
This description is normally deduced from an equivalent electrical circuit.
For instance, in \emph{Power Flow Study} (\PFS) \cite{J:PFS:1974:Stott,B:PS:2010:Milano,J:PFS:1990:Monticelli}, \emph{State Estimation} (\SE) \cite{J:PSSE:1990:Wu,B:PSSE:1999:Monticelli,B:PSSE:2004:Abur}, and \emph{Voltage Stability Assessment} (\VSA) \cite{B:PSSA:1994:Kundur,J:PSSA:1995:Kimbark,B:PSSA:1998:VanCutsem}, the use of \emph{admittance parameters} (i.e., the \emph{nodal admittance matrix}) or \emph{hybrid parameters} (i.e., \emph{hybrid parameters matrices}) is a common practice.
As the solution methods employed for these applications are computationally heavy (e.g., \cite{J:PFS:1992:Ajjarapu,J:PSSE:1971:Mendel,J:PSSE:2017:Kettner,J:PSSA:1993:Loef,J:PSSA:1997:Irisarri}), network reduction techniques, such as \emph{Kron reduction} \cite{B:CT:1959:Kron}, are often applied in order to reduce the problem size.
Thereby, the computational burden is decreased, and the execution speed is increased without the use of high-performance computers (e.g., \cite{B:HPC:2013:Khaitan}).
However, neither the reducibility of the nodal admittance matrix nor the existence of hybrid parameters are guaranteed a priori.
In order for this to be the case, the nodal admittance matrix has to satisfy certain properties (i.e., the corresponding diagonal subblocks have to be invertible).


Interestingly, most researchers and practitioners apparently ignore this fact.
There exist a handful of publications which investigate the feasibility of Kron reduction (e.g., \cite{J:CT:KR:2013:Doerfler}) and the existence of hybrid parameters matrices (e.g., \cite{J:CT:HM:1965:So,J:CT:HM:1965:Zuidweg,J:CT:HM:1966:Anderson}), but their validity is limited.
The former base their reasoning upon trivial cases (i.e., purely resistive/inductive monophase grids), and the latter establish feeble guarantees (i.e., hybrid parameters may exist for solely one partition of the nodes).
Other works, which deal with triphase power flow, show that a subblock of the nodal admittance matrix, which is obtained by removing the rows and columns associated with one single node (i.e., the slack node), has full rank in practice \cite{J:CT:YM:2017:Wang,J:CT:YM:2018:Bazrafshan}.
However, this finding cannot be generalized straightforwardly (i.e., for generic polyphase grids, or removal of several nodes).


Recently, the authors of this paper proved stronger properties for Kron reduction and hybrid parameters for monophase grids (see \cite{J:CT:YM:2017:Kettner}).
Two conditions were used in order to prove these properties: i) the connectivity of the network graph, and ii) the lossiness of the branch impedances.
If these conditions are satisfied, then Kron reduction can be performed for any set of zero-injection nodes, and a hybrid parameters matrix can be constructed for any partition of the nodes.
The theorems proven in \cite{J:CT:YM:2017:Kettner} only apply to monophase grids, and polyphase grids that can be decomposed into decoupled monophase grids using the method of \emph{symmetrical components} \cite{J:CT:HPN:1918:Fortescue}.
As known from power system analysis, the sequence decomposition only works for balanced triphase grids (i.e., grids composed of elements whose impedance/admittance matrices are \emph{symmetric circulant}), because they can be reduced to equivalent positive-sequence networks.
Therefore, the generic case of unbalanced polyphase grids cannot be treated.


This paper develops the theory for the generic case, namely unbalanced polyphase grids.
Since the method of symmetrical components cannot be applied, the grid has to be represented by a polyphase circuit, so that the electromagnetic coupling in between the phases can be accounted for properly.
Therefore, the generalization to the polyphase case is actually non-trivial.
More precisely, the electromagnetic coupling is modeled using \emph{compound electrical parameters} \cite{B:PS:1990:Arrillaga}.
It is argued that physical electrical components are represented by polyphase two-port equivalent circuits, whose compound electrical parameters are symmetric, invertible, and passive.
Via mathematical derivation and physical reasoning, it is proven that the diagonal subblocks of the compound nodal admittance matrix have full rank if the network graph is weakly connected.
Using this property, it is shown that the feasibility of Kron reduction and the existence of hybrid parameters are guaranteed under practical conditions.
In that sense, this paper provides -- for the first time in the literature -- a rigorous theoretical foundation for the analysis of polyphase power systems, in particular for applications like \PFS, \SE, and \VSA (cf. \cite{J:PFS:1982:Tiwari:1,J:PFS:1982:Tiwari:2,J:PSSE:1996:Meliopoulos}).


The remainder of this paper is structured as follows:
First, the basic theoretical foundations are laid in Sec.~\ref{Sec:Foundations}.
Thereupon, the properties of the compound nodal admittance matrix are developed in Sec.~\ref{Sec:Properties}.
Afterwards, the implications with respect to Kron reduction and hybrid parameters matrices are deduced in Sec.~\ref{Sec:Implications}.
Finally, the conclusions are drawn in Sec.~\ref{Sec:Conclusions}.

%% file: Sections/Foundations.tex
ƒmultip\WhiteSpaceMagic{-0.2cm}

\section{Foundations}
\label{Sec:Foundations}


\subsection{Numbers}

\emph{Scalars} are denoted by ordinary letters.
The \emph{real} and \emph{imaginary part} of a \emph{complex} scalar $z\in\mathbb{C}$ are denoted by $\Re\{z\}$ and $\Im\{z\}$, respectively.
Thus, $z$ can be expressed in \emph{rectangular} coordinates as $z=\Re\{z\}+j\Im\{z\}$.
The \emph{complex conjugate} of $z$ is denoted by $\Conjugate{z}$.
The \emph{absolute value} and the \emph{argument} of $z$ are denoted by $\Absolute{z}$ and $ \Argument{z}$, respectively.
Thus, $z$ can be expressed in \emph{polar} coordinates as $z=\Absolute{z}\angle\Argument{z}$.


\subsection{Set Theory}

\emph{Sets} are denoted by calligraphic letters.
The \emph{cardinality} of a set $\Set{A}$ is denoted by $\Cardinality{\Set{A}}$.
The \emph{(set-theoretic) difference} $\Set{A}\setminus\Set{B}$ of two sets $\Set{A}$ and $\Set{B}$ is defined as
\begin{equation}
	\Set{A}\setminus\Set{B} \coloneqq \left\{x\,|\,x\in\Set{A},x\notin\Set{B}\right\}
	\label{Eq:Set:Difference}
\end{equation}
The \emph{Cartesian product} $\Cartesian{\Set{A}}{\Set{B}}$ of $\Set{A}$ and $\Set{B}$ is defined as
\begin{equation}
	\Cartesian{\Set{A}}{\Set{B}} \coloneqq \left\{(a,b)\,|\,a\in\Set{A},b\in\Set{B}\right\}
	\label{Eq:Set:Cartesian}
\end{equation}
A \emph{partition} of a set $\Set{A}$ is a family of sets $\left\{\Set{A}_{k}\,|\,k\in\Set{K}\right\}$, where $\Set{K}\coloneqq\{1,\cdots,\Cardinality{\Set{K}}\}$ is an integer interval, for which
\begin{align}
	\Set{A}_{k}
	&\subseteq	\Set{A} \quad \forall k\in\Set{K}
	\label{Eq:Set:Partition:Subset}
	\\
	\Set{A}_{k}
	&\neq	\emptyset \quad \forall k\in\Set{K}
	\label{Eq:Set:Partition:Nonempty}
	\\
	\Set{A}_{k}\cap\Set{A}_{l}
	&=	\emptyset \quad \forall k,l\in\Set{K},k\neq l
	\label{Eq:Set:Partition:Disjoint}
	\\
	\bigcup\limits_{k\in\Set{K}}\Set{A}_{k}
	&=	\Set{A}
	\label{Eq:Set:Partition:Exhaustive}
\end{align}
That is, the \emph{parts} $\Set{A}_{k}$ are \emph{non-empty} and \emph{disjoint} subsets of $\Set{A}$, whose union is \emph{exhaustive}.
If $\Set{A}_{k} \subsetneq \Set{A}$ ($\forall k\in\Set{K}$), which means that $\{\Set{A}_{k}\,|\,k\in\Set{K}\}\neq\{\Set{A}\}$, the partition is \emph{non-trivial}.


\subsection{Linear Algebra}

\emph{Matrices} and \emph{vectors} are denoted by bold letters.
Consider a matrix $\Tensor{M}=(M_{rc})$ ($r\in\Set{R}$, $c\in\Set{C}$), where $\Set{R}\coloneqq\{1,\cdots,\Cardinality{\Set{R}}\}$ and $\Set{C}\coloneqq\{1,\cdots,\Cardinality{\Set{C}}\}$ are the sets of row and column indices.
The \emph{transpose} of $\Tensor{M}$ is denoted by $\Transpose{\Tensor{M}}$.
If $\Tensor{M}=\Transpose{\Tensor{M}}$, then $\Tensor{M}$ is called \emph{symmetric}.
The \emph{rank} of a matrix is the dimensionality of the vector spaces spanned by its row or column vectors.
The following properties hold
\begin{Lemma}
	\label{Lem:Rank:1}
	$\Rank(\Transpose{\Tensor{M}}\Tensor{M})=\Rank(\Tensor{M})$ $\forall\Tensor{M}\in\Field{C}^{\Cardinality{\Set{R}}\times\Cardinality{\Set{C}}}$.
\end{Lemma}
\begin{Lemma}
	\label{Lem:Rank:2}
	Let the matrices $\Tensor{A}\in\Field{C}^{\Cardinality{\Set{R}}\times\Cardinality{\Set{R}}}$ and $\Tensor{B}\in\Field{C}^{\Cardinality{\Set{C}}\times\Cardinality{\Set{C}}}$ be non-singular.
	Then, for any matrix $\Tensor{M}\in\Field{C}^{\Cardinality{\Set{R}}\times\Cardinality{\Set{C}}}$, it holds that $\Rank(\Tensor{A}\Tensor{M}) = \Rank(\Tensor{M}) = \Rank(\Tensor{M}\Tensor{B})$.
\end{Lemma}
A non-singular complex matrix is called \emph{unitary} if its inverse equals its conjugate transpose, that is $\Tensor{M}^{-1} = \Transpose{(\Conjugate{\Tensor{M}})}$.
\begin{Lemma}
	\label{Lem:Factorization}
	Let $\Tensor{M}\in\Field{C}^{\Cardinality{\Set{R}}\times\Cardinality{\Set{R}}}$ and $\Tensor{M}=\Transpose{\Tensor{M}}$.
	Then, $\Tensor{M}$ can be factorized as $\Tensor{M} = \Tensor{U}^{T}\Tensor{D}\Tensor{U}$, where $\Tensor{U}\in\Field{C}^{\Cardinality{\Set{R}}\times\Cardinality{\Set{R}}}$ is unitary, and $\Tensor{D}\in\Field{R}^{\Cardinality{\Set{R}}\times\Cardinality{\Set{R}}}$ is non-negative diagonal (\emph{Autonne-Takagi factorization}, see \cite{B:LA:2013:Horn}).
	If $\Tensor{M}$ is non-singular, then $\Tensor{D}$ is positive diagonal.
\end{Lemma}
A real symmetric matrix is called \emph{positive definite} ($\Tensor{M}\succ0$) or \emph{negative definite} ($\Tensor{M}\prec0$), respectively, when
\begin{align}
	\Tensor{M}\succ0:~		
	&	\Transpose{\Tensor{x}}\Tensor{M}\Tensor{x}>0~\forall\Tensor{x}\neq\Tensor{0}
	\\
	\Tensor{M}\prec0:~		
	&	\Transpose{\Tensor{x}}\Tensor{M}\Tensor{x}<0~\forall\Tensor{x}\neq\Tensor{0}
\end{align}
If the inequality is not strict, $\Tensor{M}$ is called \emph{positive semi-definite} ($\Tensor{M}\succeq0$) or \emph{negative semi-definite} ($\Tensor{M}\preceq0$).
\begin{Lemma}
	\label{Lem:Definite:Real}
	Let $\Tensor{M}\in\Field{C}^{\Cardinality{\Set{R}}\times\Cardinality{\Set{R}}}$ and $\Re\{\Tensor{M}\}\succ0$.
	Then, $\Tensor{M}$ is non-singular and $\Re\{\Tensor{M}^{-1}\}\succ0$ (for proof, see \cite{J:LA:1991:London}).
\end{Lemma}
\begin{Lemma}
	\label{Lem:Definite:Imaginary}
	If $\Tensor{M}\in\Field{C}^{\Cardinality{\Set{R}}\times\Cardinality{\Set{R}}}$ and $\Im\{\Tensor{M}\}\succ0$.
	Then, $\Tensor{M}$ is non-singular and $\Im\{\Tensor{M}^{-1}\}\prec0$ (for proof, see \cite{J:LA:1972:Fan}).
\end{Lemma}
Let $\left\{\Set{R}_{i}\,|\,i\in\Set{I}\right\}$ and $\left\{\Set{C}_{j}\,|\,j\in\Set{J}\right\}$ be partitions of $\Set{R}$ and $\Set{C}$, where $\Set{I}\coloneqq\{1,\cdots,\Cardinality{\Set{I}}\}$ and $\Set{J}\coloneqq\{1,\cdots,\Cardinality{\Set{J}}\}$.
The block formed by the intersection of the rows $\Set{R}_{i}$ ($i\in\Set{I}$) with the columns $\Set{C}_{j}$ ($j\in\Set{J}$) is denoted by $\Tensor{M}_{ij}$.
That is, $\Tensor{M}=\left(\Tensor{M}_{ij}\right)$.
Let $\Tensor{M}_{\Set{I}'\times\Set{J}'}$ ($\Set{I}'\subseteq\Set{I}$, $\Set{J}'\subseteq\Set{J}$) be the submatrix consisting of the blocks $\Tensor{M}_{ij}$ ($i\in\Set{I}'$, $j\in\Set{J}'$).
Now, consider the particular case $\Set{I}'\coloneqq\Set{I}\setminus\{\Cardinality{\Set{I}}\}$ and $\Set{J}'\coloneqq\Set{J}\setminus\{\Cardinality{\Set{J}}\}$, and define
\begin{equation}
	\Tensor{M}
	=	\left[
		\begin{array}{cc}
			\Tensor{A}	&\Tensor{B}	\\
			\Tensor{C}	&\Tensor{D}	
		\end{array}
		\right]
	\coloneqq	\left[
		\begin{array}{ll}
			\Tensor{M}_{\Set{I}'\times\Set{J}'}							&\Tensor{M}_{\Set{I}'\times\{\Cardinality{\Set{J}}\}}	\\
			\Tensor{M}_{\{\Cardinality{\Set{I}}\}\times\Set{J}'}	&\Tensor{M}_{\Cardinality{\Set{I}}\Cardinality{\Set{J}}}
		\end{array}
		\right]
\end{equation}
If $\Tensor{D}$ is invertible, the \emph{Schur complement} $\Schur{\Tensor{M}}{\Tensor{D}}$ of $\Tensor{D}$ in $\Tensor{M}$ is
\begin{equation}
	\Schur{\Tensor{M}}{\Tensor{D}} \coloneqq \Tensor{A}-\Tensor{B}\Tensor{D}^{-1}\Tensor{C}
	\label{Eq:Schur}
\end{equation}
The following properties hold (see \cite{B:LA:2006:Zhang}).
\begin{Lemma}
	\label{Lem:Schur:Determinant}
	$\det(\Tensor{M})=\det(\Tensor{D})\det(\Schur{\Tensor{M}}{\Tensor{D}})$.
\end{Lemma}
\begin{Lemma}
	\label{Lem:Schur:Blockform}
	Let $i\in\Set{I}'$ and $j\in\Set{J}'$.
	Then
	\begin{equation}
		(\Schur{\Tensor{M}}{\Tensor{D}})_{ij}
		=	\Tensor{A}_{ij} - \Tensor{B}_{i}\Tensor{D}^{-1}\Tensor{C}_{j}
		=	\left[
			\begin{array}{ll}
				\Tensor{A}_{ij}	&\Tensor{B}_{i}\\
				\Tensor{C}_{j}		&\Tensor{D}
			\end{array}
			\right]
			/~
			\Tensor{D}
	\end{equation}
\end{Lemma}
The \emph{Kronecker product} $\Kronecker{\Tensor{A}}{\Tensor{B}}$ of two generic matrices $\Tensor{A}$ and $\Tensor{B}$ is a block matrix, whose blocks $(\Kronecker{\Tensor{A}}{\Tensor{B}})_{ij}$ are defined as the product of the corresponding element $A_{ij}$ of $\Tensor{A}$ and $\Tensor{B}$.
\begin{equation}
	\Kronecker{\Tensor{A}}{\Tensor{B}}:
	(\Kronecker{\Tensor{A}}{\Tensor{B}})_{ij} = A_{ij}\Tensor{B}
\end{equation}
The following property holds (see \cite{B:LA:1981:Graham}).
\begin{Lemma}
	\label{Lem:Kronecker:Rank}
	$\Rank(\Kronecker{\Tensor{A}}{\Tensor{B}})=\Rank(\Tensor{A})\cdot\Rank(\Tensor{B})$.
\end{Lemma}


\subsection{Graph Theory}

A \emph{directed graph} $\Graph{G}=(\Set{V},\Set{E})$ consists of a set of \emph{vertices} $\Vertices{\Graph{G}}\coloneqq\Set{V}$ and a set of \emph{directed edges} $\Edges{\Graph{G}}\coloneqq\Set{E}$, where
\begin{equation}
	\Edges{\Graph{G}}\subseteq\{(v,w)\in\Cartesian{\Vertices{\Graph{G}}}{\Vertices{\Graph{G}}}\,|\,v\neq w\}
\end{equation}
The sets of \emph{outgoing} and \emph{incoming edges} of a vertex $v\in\Vertices{\Graph{G}}$ are denoted by $\Outgoing{\Graph{G}}{v}$ and $\Incoming{\Graph{G}}{v}$.
Formally
\begin{align}
	\Outgoing{\Graph{G}}{v}
	&\coloneqq	\{ e\in\Edges{\Graph{G}} \mid e=(v,w),\,w\in\Vertices{\Graph{G}} \}
	\label{Eq:Edge:Outgoing}
	\\
	\Incoming{\Graph{G}}{v}
	&\coloneqq	\{ e\in\Edges{\Graph{G}} \mid e=(w,v),\,w\in\Vertices{\Graph{G}} \}
	\label{Eq:Edge:Incoming}
\end{align}
A set of \emph{internal edges} $\Internal{\Graph{G}}{\Set{W}}$ with respect to $\Set{W}\subseteq\Vertices{\Graph{G}}$ contains all directed edges that start and end in $\Set{W}$.
So
\begin{equation}
	\Internal{\Graph{G}}{\Set{W}}
	\coloneqq	\{(u,v)\in\Edges{\Graph{G}} \mid u,v\in\Set{W}\subseteq\Vertices{\Graph{G}}\}
	\label{Eq:Edge:Internal}
\end{equation}
A \emph{cut-set} $\Cut{\Graph{G}}{\Set{W}}$ with respect to $\Set{W}\subsetneq\Vertices{\Graph{G}}$ contains all directed edges that start in $\Set{W}$ and end in $\Vertices{\Graph{G}}\setminus\Set{W}$.
That is
\begin{equation}
	\Cut{\Graph{G}}{\Set{W}}
	\coloneqq
		\left\{
			(u,v)\in\Vertices{\Graph{G}}
			\left|~
				\begin{aligned}
					u&\in\Set{W}\subsetneq\Vertices{\Graph{G}},\\
					v&\in\Vertices{\Graph{G}}\setminus\Set{W}
				\end{aligned}
			\right.
		\right\}
	\label{Eq:Edge:Cut}
\end{equation}
The connectivity of the graph $\Graph{G}$ is given by its \emph{(edge-to-vertex) incidence matrix} $\Tensor{A}_{\Graph{G}}$, whose elements are given by (see \cite{B:GT:2013:Gross})
\begin{equation}
	A_{\Graph{G},iv}
	\coloneqq
		\left\{
		\begin{array}{rl}
			+1	&\text{if}~e_{i}\in\Outgoing{\Graph{G}}{v}\\
			-1	&\text{if}~e_{i}\in\Incoming{\Graph{G}}{v}\\
			0	&\text{otherwise}
		\end{array}
		\right.
	\label{Eq:Graph:Indicence}
\end{equation}
\noindent
A directed graph $\Graph{G}$ is said to be \emph{weakly connected} if there exists a connecting path (which need not respect the directivity of the edges) between any pair of vertices.
\begin{Lemma}
	\label{Lem:Incidence:Rank}
	If the directed graph $\Graph{G}$ is weakly connected, then $\Rank(\Tensor{A}_{\Graph{G}})=\Cardinality{\Vertices{\Graph{G}}}-1$ (for proof, see \cite{B:CT:1969:Desoer}).
\end{Lemma}


\subsection{Power System Analysis}

\begin{figure}
	\centering
	
	\subfloat[$\Pie$-section equivalent circuit.]
	{%
		\centering
		\input{"Figures/Equivalent_Pie_Section"}
		\label{Fig:Component:Pie}
	}
	
	\subfloat[$\Tea$-section equivalent circuit.]
	{%
		\centering
		\input{"Figures/Equivalent_Tea_Section"}
		\label{Fig:Component:Tea}
	}
	
	\caption{Polyphase two-port equivalent circuits of the components of the grid.}
	\label{Fig:Component}
\end{figure}
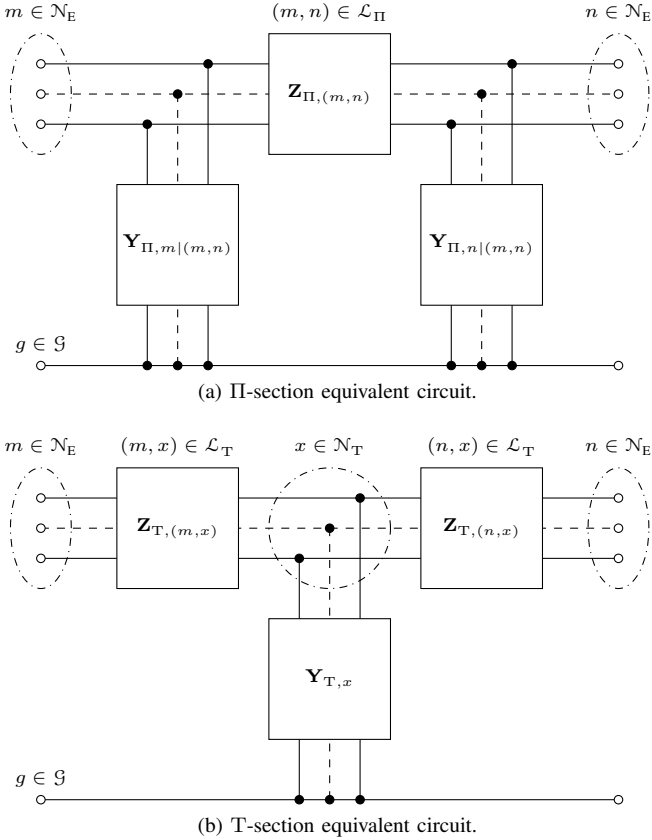


Consider the case of an unbalanced polyphase power system, which is equipped with a neutral conductor.
With regard to the wiring, the following is assumed:
\begin{Hypothesis}
	\label{Hyp:Neutral}
	The reference point of every voltage or current source is connected with the neutral conductor.
	Furthermore, the neutral conductor is grounded using an effective earthing system, which is capable of establishing a null voltage between the neutral conductor and the physical ground (see \cite{J:PSA:1998:Das,J:PFS:2011:Chen}).
\end{Hypothesis}
\noindent
Therefore, the phase-to-neutral voltages effectively correspond to phase-to-ground voltages.
Let $\Set{G}\coloneqq\{0\}$ be the \emph{ground node}, and $\Set{P}\coloneqq\{1,\ldots,\Cardinality{\Set{P}}\}$ the \emph{phases}.
An array of terminals which belong together (one for every phase $p\in\Set{P}$) form a \emph{polyphase node}.
Define $\Set{N}_{\mathrm{E}}\coloneqq\{1,\ldots,\Cardinality{\Set{N}_{\mathrm{E}}}\}$ as the set of the \emph{physically existent} polyphase nodes of the grid (i.e., where actual voltages and currents could be measured).
The grid consists of electrical components that link the polyphase nodes with each other and the ground.
As to the grid, the following is presumed:
\begin{Hypothesis}
	\label{Hyp:Component}
	The grid consists of electrical components that are linear and passive.
	Further, electromagnetic coupling only matters inside of electrical components, but not between them.
	Hence, in a per-unit grid model, they are represented either by $\Pie$-section or $\Tea$-section polyphase two-port equivalent circuits without mutual coupling (see Fig.~\ref{Fig:Component} and \cite{B:PS:1990:Arrillaga}).
\end{Hypothesis}


Every $\Tea$-section equivalent circuit comes with an additional polyphase node (see Fig.~\ref{Fig:Component:Tea}).
These nodes are purely \emph{virtual}.
That is, they are part of the model, but do not exist in reality.
Let $\Set{N}_{\Tea}$ encompass all virtual polyphase nodes originating from the $\Tea$-section equivalent circuits.
The topology of the grid model is described by the directed graph $\Graph{G}=\left(\Set{V},\Set{E}\right)$, which is constructed as follows.
Define $\Set{N}\coloneqq\Set{N}_{\mathrm{E}}\cup\Set{N}_{\Tea}$ as the set of all polyphase nodes (i.e., physical and virtual).
The vertices are
\begin{equation}
	\Set{V} \coloneqq \Set{N} \cup \Set{G}
\end{equation}
The edges fall into two categories, namely \emph{polyphase branches} and \emph{polyphase shunts}.
The former connect a pair of polyphase nodes, the latter a polyphase node and ground.
Let $\Set{L}_{\Pie}\subseteq\Set{N}\times\Set{N}$ and $\Set{L}_{\Tea}\subseteq\Set{N}\times\Set{N}_{T}$ be the polyphase branches associated with the $\Pie$-section and $\Tea$-section equivalent circuits, respectively.
The set of all polyphase branches is $\Set{L}\coloneqq\Set{L}_{\Pie}\cup\Set{L}_{\Tea}$.
Similarly, let $\Set{T}_{\mathrm{E}}\coloneqq\Set{N}_{\mathrm{E}}\times\Set{G}$ and $\Set{T}_{\Tea}\coloneqq\Set{N}_{\Tea}\times\Set{G}$ be the polyphase shunts associated with $\Set{N}_{\mathrm{E}}$ and $\Set{N}_{\Tea}$, respectively.
Thus, the set of all polyphase shunts is $\Set{T}\coloneqq\Set{T}_{\mathrm{E}}\cup\Set{T}_{\Tea}$.
The edges are obtained as
\begin{equation}
	\Set{E} \coloneqq \Set{L} \cup \Set{T}
\end{equation}	


The polyphase branches are related to the \emph{longitudinal} electrical parameters of the polyphase two-port equivalents.
More precisely, every polyphase branch $\ell\in\Set{L}$ is associated with a \emph{compound branch impedance $\Tensor{Z}_{\ell}$}, which is given by
\begin{equation}
	\Tensor{Z}_{\ell}
	\coloneqq
		\left\{
		\begin{array}{cl}
			\Tensor{Z}_{\Pie,(m,n)}	&\text{if}~\ell=(m,n)\in\Set{L}_{\Pie}\\
			\Tensor{Z}_{\Tea,(n,x)}	&\text{if}~\ell=(n,x)\in\Set{L}_{\Tea}
		\end{array}
		\right.
	\label{Eq:Branch:Impedance}
\end{equation}
Similarly, the polyphase shunts are related to the \emph{transversal} electrical parameters of the polyphase two-port equivalents.
The aggregated shunt admittance $\Tensor{Y}_{\Pie,n}$ resulting from the $\Pie$-section equivalent circuits connected to the polyphase node $n\in\Set{N}_{\text{E}}$ is given by
\begin{equation}
	\Tensor{Y}_{\Pie,n}
	\coloneqq		\sum\limits_{(n,m)\in\Set{L}_{\Pie}}\Tensor{Y}_{\Pie,n|(n,m)}
					+	\sum\limits_{(m,n)\in\Set{L}_{\Pie}}\Tensor{Y}_{\Pie,n|(m,n)}
	\label{Eq:Shunt:Aggregate}
\end{equation}
Accordingly, the \emph{compound shunt admittance} $\Tensor{Y}_{t}$ associated with a polyphase shunt $t\in\Set{T}$ is given by
\begin{equation}
	\Tensor{Y}_{t}
	\coloneqq
		\left\{
		\begin{array}{cl}
			\Tensor{Y}_{\Pie,n}	&\text{if}~t=(n,g)\in\Set{T}_{\mathrm{E}}\\
			\Tensor{Y}_{\Tea,x}	&\text{if}~t=(x,g)\in\Set{T}_{\Tea}
		\end{array}
		\right.
	\label{Eq:Shunt:Admittance}
\end{equation}
With respect to the compound electrical parameters of the grid model, the following assumption is made:
\begin{Hypothesis}
	\label{Hyp:Parameter}
	The compound branch impedances $\Tensor{Z}_{\ell}$ defined by \eqref{Eq:Branch:Impedance} are symmetric, invertible, and passive.
	That is
	\begin{equation}
		\forall\ell\in\Set{L}:
		~
		\left[~
		\begin{aligned}
			\Tensor{Z}_{\ell}
			&=	\Transpose{\Tensor{Z}_{\ell}}
			\\
			\exists\Tensor{Y}_{\ell}
			&=	\Inverse{\Tensor{Z}_{\ell}}
			\\
			\Re\{\Tensor{Z}_{\ell}\}
			&\succeq	0
		\end{aligned}
		\right.
		\label{Eq:Parameter:Branch}
	\end{equation}
	In particular, this implies $\Tensor{Z}_{\ell}\neq\Tensor{0}$.
	Conversely, the compound shunt admittances $\Tensor{Y}_{t}$ defined by \eqref{Eq:Shunt:Admittance} may be zero.
	If not, then they are also symmetric, invertible, and passive.
	That is
	\begin{equation}
		t\in\Set{T}~\text{with}~\Tensor{Y}_{t}\neq\Tensor{0}:
		~
		\left[~
		\begin{aligned}
			\Tensor{Y}_{t}
			&=	\Transpose{\Tensor{Y}_{t}}
			\\
			\exists\Tensor{Z}_{t}
			&=	\Inverse{\Tensor{Y}_{t}}
			\\
			\Re\{\Tensor{Y}_{t}\}
			&\succeq	\Tensor{0}
		\end{aligned}
		\right.
	\label{Eq:Parameter:Shunt}
	\end{equation}
\end{Hypothesis}
\noindent
In practice, the above-stated assumptions are valid for a broad variety of power system components, like transmission lines, transformers, and various \FACTS devices.
Further information about this subject is given in App.~{App:Equipment}.


\begin{figure}
	\centering
	\input{"Figures/Node_Quantities"}
	\caption{Definition of the injected currents, nodal voltages, compound branch impedances, and compound shunt admittances.}
	\label{Fig:Node}
\end{figure}
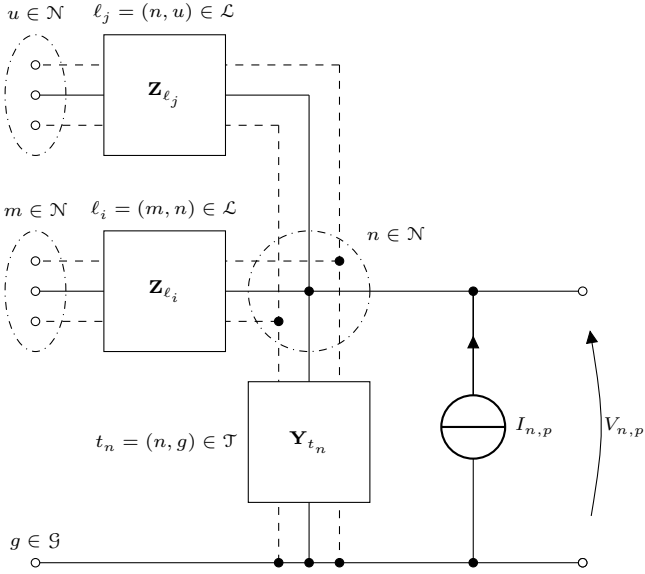

Let $V_{n,p}$ and $I_{n,p}$ be the phasors of the nodal voltage and the injected current in phase $p\in\Set{P}$ of the polyphase node $n\in\Set{N}$.
By definition, the nodal voltage is referenced to the ground node, and the injected current flows from the ground node into the corresponding terminal (see Fig.~\ref{Fig:Node}).
Analogous quantities are defined for a polyphase node as a whole
\begin{align}
	\Tensor{V}_{n}
	&\coloneqq	\Column_{p\in\Set{P}}(V_{n,p})
	\\
	\Tensor{I}_{n}
	&\coloneqq	\Column_{p\in\Set{P}}(I_{n,p})
\end{align}
and for the grid as a whole
\begin{align}
	\Tensor{V}
	&\coloneqq	\Column_{n\in\Set{N}}(\Tensor{V}_{n})
	\label{Eq:Node:Voltage}
	\\
	\Tensor{I}
	&\coloneqq	\Column_{n\in\Set{N}}(\Tensor{I}_{n})
	\label{Eq:Node:Current}
\end{align}
where the operator $\Column$ constructs a (block) column vector.


The \emph{primitive compound branch admittance matrix $\Tensor{Y}_{\Set{L}}$} and the \emph{primitive compound shunt admittance matrix} $\Tensor{Y}_{\Set{T}}$ are
\begin{align}
	\Tensor{Y}_{\Set{L}}
	&\coloneqq	\Diagonal_{\ell\in\Set{L}}(\Tensor{Y}_{\ell})
	\label{Eq:Branch:Admittance:Primitive}
	\\
	\Tensor{Y}_{\Set{T}}
	&\coloneqq	\Diagonal_{t\in\Set{T}}(\Tensor{Y}_{t})
	\label{Eq:Shunt:Admittance:Primitive}
\end{align}
where the operator $\Diagonal$ constructs a (block) diagonal matrix.
Let $\Graph{B}\coloneqq(\Set{N},\Set{L})$ represent the subgraph of $\Graph{G}$ comprising the branches only.
Define its \emph{polyphase incidence matrix} $\Tensor{A}_{\Graph{B}}^{\Set{P}}$ as
\begin{equation}
	\Tensor{A}_{\Graph{B}}^{\Set{P}}
	\coloneqq	\Kronecker{\Tensor{A}_{\Graph{B}}}{\Diagonal(\Tensor{1}_{\Cardinality{\Set{P}}})}
	\label{Eq:Incidence:Polyphase}
\end{equation}
where $\Tensor{1}_{\Cardinality{\Set{P}}}$ is a vector of ones with length $\Cardinality{\Set{P}}$.
The \emph{compound nodal admittance matrix} $\Tensor{Y}$, which relates $\Tensor{I}$ with $\Tensor{V}$ via
\begin{equation}
	\Tensor{I} = \Tensor{Y}\Tensor{V}
	\label{Eq:Ohm}
\end{equation}
(i.e., \emph{Ohm's law}) is given by (see \cite{B:PS:1990:Arrillaga,B:CT:1969:Desoer})
\begin{equation}
	\Tensor{Y}
	=		\Transpose{(\Tensor{A}_{\Graph{B}}^{\Set{P}})}\Tensor{Y}_{\Set{L}}\Tensor{A}_{\Graph{B}}^{\Set{P}}
		+	\Tensor{Y}_{\Set{T}}
	\label{Eq:Node:Admittance}
\end{equation}
By definition \eqref{Eq:Node:Voltage}--\eqref{Eq:Node:Current}, the vectors $\Tensor{V}$ and $\Tensor{I}$ are composed of blocks which correspond to the polyphase nodes of the grid.
Therefore, $\Tensor{Y}$ can be written in a block form as $\Tensor{Y}=(\Tensor{Y}_{mn})$, where $\Tensor{Y}_{mn}$ relates $\Tensor{I}_{m}$ with $\Tensor{V}_{n}$ ($m,n\in\Set{N}$).
As known from circuit theory, it holds that (see \cite{B:CT:1969:Desoer})
\begin{Lemma}
	\label{Lem:Admittance:Sum}
	$t=(n,g)\in\Set{T}$:~$\sum\limits_{m\in\Set{N}}\Tensor{Y}_{nm} = \sum\limits_{m\in\Set{N}}\Tensor{Y}_{mn} = \Tensor{Y}_{t}$.
\end{Lemma}


%% file: Figures/Equivalent_Pie_Section.tex
\begin{circuitikz}
	\scriptsize
	
	
	\def\ShuntPosition{2.0}
	\def\ShuntHeight{1.6}
	\def\PortPosition{3.8}
	\def\PhaseHeight{3.6}
	\def\BlockSize{1.6}
	\def\WireSpacing{0.4}
	
	
	
	\coordinate (PhaseLeft) at (-\PortPosition,\PhaseHeight);
	\coordinate (PhaseRight) at (\PortPosition,\PhaseHeight);
	\coordinate (NeutralLeft) at (-\PortPosition,0);
	\coordinate (NeutralRight) at (\PortPosition,0);
	
	\draw[dashed] (PhaseLeft) to[short] (PhaseRight);
	\draw (PhaseLeft) to[open,o-o] (PhaseRight);
	\draw ($(PhaseLeft)+\WireSpacing*(0,1)$) to [short,o-o] ($(PhaseRight)+\WireSpacing*(0,1)$);
	\draw ($(PhaseLeft)-\WireSpacing*(0,1)$) to [short,o-o] ($(PhaseRight)-\WireSpacing*(0,1)$);
	\draw (NeutralLeft) to[short,o-o] (NeutralRight);
	
	\coordinate (Branch) at (0,\PhaseHeight);
	\draw[fill=white] ($(Branch)-0.5*\BlockSize*(1,1)$) rectangle ($(Branch)+0.5*\BlockSize*(1,1)$);
	\node at (Branch) {$\Tensor{Z}_{\Pie,(m,n)}$};
	\node at ($(Branch)+2.7*(0,\WireSpacing)$) {$(m,n)\in\Set{L}_{\Pie}$};
	
	\draw[dashdotted] (PhaseLeft) ellipse (0.40 and 0.80);
	\node at ($(PhaseLeft)+2.7*(0,\WireSpacing)$) {$m\in\Set{N}_{\text{E}}$};
	
	\draw[dashdotted] (PhaseRight) ellipse (0.40 and 0.80);	
	\node at ($(PhaseRight)+2.7*(0,\WireSpacing)$) {$n\in\Set{N}_{\text{E}}$};
	
	\node at ($(NeutralLeft)+0.7*(0,\WireSpacing)$) {$g\in\Set{G}$};
	
	
	
	\coordinate (ShuntLeftPhase) at (-\ShuntPosition,\PhaseHeight);
	\coordinate (ShuntLeftNeutral) at (-\ShuntPosition,0);
	
	\draw[dashed] (ShuntLeftPhase) to[short] (ShuntLeftNeutral);
	\draw (ShuntLeftPhase) to[open,*-*] (ShuntLeftNeutral);
	\draw ($(ShuntLeftPhase)+\WireSpacing*(1,1)$) to [short,*-*] ($(ShuntLeftNeutral)+\WireSpacing*(1,0)$);
	\draw ($(ShuntLeftPhase)-\WireSpacing*(1,1)$) to [short,*-*] ($(ShuntLeftNeutral)-\WireSpacing*(1,0)$);
	
	\coordinate (ShuntLeft) at (-\ShuntPosition,\ShuntHeight);
	\draw[fill=white] ($(ShuntLeft)-0.5*\BlockSize*(1,1)$) rectangle ($(ShuntLeft)+0.5*\BlockSize*(1,1)$);
	\node at (ShuntLeft) {$\Tensor{Y}_{\Pie,m|(m,n)}$};
	
	
	
	\coordinate (ShuntRightPhase) at (\ShuntPosition,\PhaseHeight);
	\coordinate (ShuntRightNeutral) at (\ShuntPosition,0);
	
	\draw[dashed] (ShuntRightPhase) to[short] (ShuntRightNeutral);
	\draw (ShuntRightPhase) to[open,*-*] (ShuntRightNeutral);
	\draw ($(ShuntRightPhase)+\WireSpacing*(1,1)$) to [short,*-*] ($(ShuntRightNeutral)+\WireSpacing*(1,0)$);
	\draw ($(ShuntRightPhase)-\WireSpacing*(1,1)$) to [short,*-*] ($(ShuntRightNeutral)-\WireSpacing*(1,0)$);
	
	\coordinate (ShuntRight) at (\ShuntPosition,\ShuntHeight);
	\draw[fill=white] ($(ShuntRight)-0.5*\BlockSize*(1,1)$) rectangle ($(ShuntRight)+0.5*\BlockSize*(1,1)$);
	\node at (ShuntRight) {$\Tensor{Y}_{\Pie,n|(m,n)}$};
		
\end{circuitikz}

%% file: Figures/Equivalent_Tea_Section.tex
\begin{circuitikz}
	\scriptsize
		
	\def\BranchPosition{2.0}
	\def\ShuntPosition{2.0}
	\def\ShuntHeight{1.6}
	\def\PortPosition{3.8}
	\def\PhaseHeight{3.6}
	\def\BlockSize{1.6}
	\def\WireSpacing{0.4}
	
	
	
	\coordinate (PhaseLeft) at (-\PortPosition,\PhaseHeight);
	\coordinate (PhaseRight) at (\PortPosition,\PhaseHeight);
	\coordinate (NeutralLeft) at (-\PortPosition,0);
	\coordinate (NeutralRight) at (\PortPosition,0);
	
	\draw[dashed] (PhaseLeft) to[short] (PhaseRight);
	\draw(PhaseLeft) to[open,o-o] (PhaseRight);
	\draw ($(PhaseLeft)+\WireSpacing*(0,1)$) to[short,o-o] ($(PhaseRight)+\WireSpacing*(0,1)$);
	\draw ($(PhaseLeft)-\WireSpacing*(0,1)$) to[short,o-o] ($(PhaseRight)-\WireSpacing*(0,1)$);
		
	
	\coordinate (BranchLeft) at (-\BranchPosition,\PhaseHeight);
	\draw[fill=white] ($(BranchLeft)-0.5*\BlockSize*(1,1)$) rectangle ($(BranchLeft)+0.5*\BlockSize*(1,1)$);
	\node at (BranchLeft) {$\Tensor{Z}_{\Tea,(m,x)}$};
	\node at ($(BranchLeft)+2.7*(0,\WireSpacing)$) {$(m,x)\in\Set{L}_{\Tea}$};
	
	\draw[dashdotted] (PhaseLeft) ellipse (0.40 and 0.80);	
	\node at ($(PhaseLeft)+2.7*(0,\WireSpacing)$) {$m\in\Set{N}_{\text{E}}$};
	
	
	\coordinate (BranchRight) at (\BranchPosition,\PhaseHeight);
	\draw[fill=white] ($(BranchRight)-0.5*\BlockSize*(1,1)$) rectangle ($(BranchRight)+0.5*\BlockSize*(1,1)$);
	\node at (BranchRight) {$\Tensor{Z}_{\Tea,(n,x)}$};
	\node at ($(BranchRight)+2.7*(0,\WireSpacing)$) {$(n,x)\in\Set{L}_{\Tea}$};
	
	\draw[dashdotted] (PhaseRight) ellipse (0.40 and 0.80);
	\node at ($(PhaseRight)+2.7*(0,\WireSpacing)$) {$n\in\Set{N}_{\text{E}}$};
	
	
	\draw (NeutralLeft) to[short,o-o] (NeutralRight);
	\node at ($(NeutralLeft)+0.7*(0,\WireSpacing)$) {$g\in\Set{G}$};
	
	
	
	\coordinate (ShuntPhase) at (0,\PhaseHeight);
	\coordinate (ShuntNeutral) at (0,0);
	
	\draw[dashed] (ShuntPhase) to[short] (ShuntNeutral);
	\draw (ShuntPhase) to[open,*-*] (ShuntNeutral);
	\draw ($(ShuntPhase)+\WireSpacing*(1,1)$) to[short,*-*] ($(ShuntNeutral)+\WireSpacing*(1,0)$);
	\draw ($(ShuntPhase)-\WireSpacing*(1,1)$) to[short,*-*] ($(ShuntNeutral)-\WireSpacing*(1,0)$);
	
	\coordinate (Shunt) at (0,\ShuntHeight);
	\draw[fill=white] ($(Shunt)-0.5*\BlockSize*(1,1)$) rectangle ($(Shunt)+0.5*\BlockSize*(1,1)$);
	\node at (Shunt) {$\Tensor{Y}_{\Tea,x}$};
	
	\draw[dashdotted] (0,\PhaseHeight) circle (2*\WireSpacing);
	\node at ($(ShuntPhase)+2.7*(0,\WireSpacing)$) {$x\in\Set{N}_{\Tea}$};
	
\end{circuitikz}

%% file: Figures/Node_Quantities.tex
\begin{circuitikz}
	\scriptsize
	
	\def\BranchPosition{1.9}	
	\def\BranchHeightLower{3.6}	
	\def\BranchHeightUpper{6.2}	
	\def\ShuntHeight{1.6} 
	\def\PortPosition{3.6}	
	\def\BlockSize{1.6}	
	\def\WireSpacing{0.4}	
	
	
	\coordinate (PhaseLeftLower) at (-\PortPosition,\BranchHeightLower);
	\coordinate (PhaseLeftUpper) at (-\PortPosition,\BranchHeightUpper);		
	\coordinate (PhaseCentreLower) at (0,\BranchHeightLower);
	\coordinate (PhaseCentreUpper) at (0,\BranchHeightUpper);
	\coordinate (PhaseRight) at (\PortPosition,\BranchHeightLower);
	
	\coordinate (NeutralLeft) at (-\PortPosition,0);	
	\coordinate (NeutralCentre) at (0,0);
	\coordinate (NeutralRight) at (\PortPosition,0);
	
	
	
	\draw (NeutralLeft) to[short,o-o] (NeutralRight);
	\node at ($(NeutralLeft)+0.7*\WireSpacing*(0,1)$) {$g\in\Set{G}$};
	
	
	
	\draw (PhaseCentreLower) to (PhaseLeftLower);	
	\draw[dashed] ($(PhaseCentreLower)+\WireSpacing*(1,1)$) to ($(PhaseLeftLower)+\WireSpacing*(0,1)$);	
	\draw[dashed] ($(PhaseCentreLower)-\WireSpacing*(1,1)$) to ($(PhaseLeftLower)-\WireSpacing*(0,1)$);
	
	\draw (PhaseCentreLower) to[open,-o] (PhaseLeftLower);
	\draw ($(PhaseCentreLower)+\WireSpacing*(1,1)$) to[open,-o] ($(PhaseLeftLower)+\WireSpacing*(0,1)$);
	\draw ($(PhaseCentreLower)-\WireSpacing*(1,1)$) to[open,-o] ($(PhaseLeftLower)-\WireSpacing*(0,1)$);
	
	\draw[dashdotted] (PhaseLeftLower) ellipse (0.40 and 0.80);
	\node at ($(PhaseLeftLower)+2.7*\WireSpacing*(0,1)$) {$m\in\Set{N}$};
	
	\coordinate (BranchLower) at (-\BranchPosition,\BranchHeightLower);
	\draw[fill=white] ($(BranchLower)-0.5*\BlockSize*(1,1)$) rectangle ($(BranchLower)+0.5*\BlockSize*(1,1)$);
	\node at (BranchLower) {$\Tensor{Z}_{\ell_{i}}$};
	
	\node at ($(BranchLower)+2.7*\WireSpacing*(0,1)$) {$\ell_{i}=(m,n)\in\Set{L}$};
	
	
	
	\draw (PhaseCentreUpper) to (PhaseLeftUpper);	
	\draw[dashed] ($(PhaseCentreUpper)+\WireSpacing*(1,1)$) to ($(PhaseLeftUpper)+\WireSpacing*(0,1)$);	
	\draw[dashed] ($(PhaseCentreUpper)-\WireSpacing*(1,1)$) to ($(PhaseLeftUpper)-\WireSpacing*(0,1)$);	
	
	\draw (PhaseCentreUpper) to[open,-o] (PhaseLeftUpper);
	\draw ($(PhaseCentreUpper)+\WireSpacing*(1,1)$) to[open,-o] ($(PhaseLeftUpper)+\WireSpacing*(0,1)$);
	\draw ($(PhaseCentreUpper)-\WireSpacing*(1,1)$) to[open,-o] ($(PhaseLeftUpper)-\WireSpacing*(0,1)$);
	\draw[dashdotted] (PhaseLeftUpper) ellipse (0.40 and 0.80);
	\node at ($(PhaseLeftUpper)+2.7*\WireSpacing*(0,1)$) {$u\in\Set{N}$};
	
	\coordinate (BranchUpper) at (-\BranchPosition,\BranchHeightUpper);
	\draw[fill=white] ($(BranchUpper)-0.5*\BlockSize*(1,1)$) rectangle ($(BranchUpper)+0.5*\BlockSize*(1,1)$);
	\node at (BranchUpper) {$\Tensor{Z}_{\ell_{j}}$};
	
	\node at ($(BranchUpper)+2.7*\WireSpacing*(0,1)$) {$\ell_{j}=(n,u)\in\Set{L}$};
	
	
	
	\draw (PhaseCentreUpper) to (NeutralCentre);	
	\draw[dashed] ($(PhaseCentreUpper)+\WireSpacing*(1,1)$) to ($(NeutralCentre)+\WireSpacing*(1,0)$);
	\draw[dashed] ($(PhaseCentreUpper)-\WireSpacing*(1,1)$) to ($(NeutralCentre)-\WireSpacing*(1,0)$);
	
	\draw (PhaseCentreLower) to[open,*-*] (NeutralCentre);	
	\draw ($(PhaseCentreLower)+\WireSpacing*(1,1)$) to[open,*-*] ($(NeutralCentre)+\WireSpacing*(1,0)$);
	\draw ($(PhaseCentreLower)-\WireSpacing*(1,1)$) to[open,*-*] ($(NeutralCentre)-\WireSpacing*(1,0)$);
	
	\coordinate (Shunt) at (0,\ShuntHeight);
	\draw[fill=white] ($(Shunt)-0.5*\BlockSize*(1,1)$) rectangle ($(Shunt)+0.5*\BlockSize*(1,1)$);
	\node at (Shunt) {$\Tensor{Y}_{t_{n}}$};
	
	\draw[dashdotted] (PhaseCentreLower) circle (2*\WireSpacing);
	\node at ($(PhaseCentreLower)+1.9*\WireSpacing*(1.5,1)$) {$n\in\Set{N}$};
	
	\node at ($(Shunt)-4.7*\WireSpacing*(1,0)$) {$t_{n}=(n,g)\in\Set{T}$};
	
	
	
	\draw (PhaseRight) to (PhaseCentreLower);
	\draw (NeutralRight) to[open,v=$V_{n,p}$,o-o] (PhaseRight);
	\draw ($(0.6*\PortPosition,0)$) to[current source,i_=$I_{n,p}$,*-*] ($(0.6*\PortPosition,\BranchHeightLower)$);
	
\end{circuitikz}

%% file: Sections/Properties.tex
\section{Properties}
\label{Sec:Properties}

The properties of the compound nodal admittance matrix are developed in two steps.
First, it is proven that the compound shunt admittances determine the rank of the whole matrix in Sec.~\ref{Sec:Properties:Rank}.
Afterwards, it is shown that the compound branch impedances determine the rank of its diagonal blocks in Sec.~\ref{Sec:Properties:Blockrank}).


\subsection{Rank}
\label{Sec:Properties:Rank}

\begin{figure}[t]
	\centering
	\subfloat[$\Tensor{Y}_{t}=\Tensor{0}~\forall t\in\Set{T}$.]
	{
		\centering
		\input{"Figures/Rank_Graph"}
		\label{Fig:Grid:Original}
	}
	\subfloat[$\exists t\in\Set{T}$ s.t. $\Tensor{Y}_{t}\neq\Tensor{0}$.]
	{
		\centering
		\input{"Figures/Rank_Graph_Augmented"}
		\label{Fig:Grid:Augmented}
	}

	\caption{Proof of Theorem \ref{Thm:Rank} (surfaces indicate weakly connected graphs).}
\end{figure}
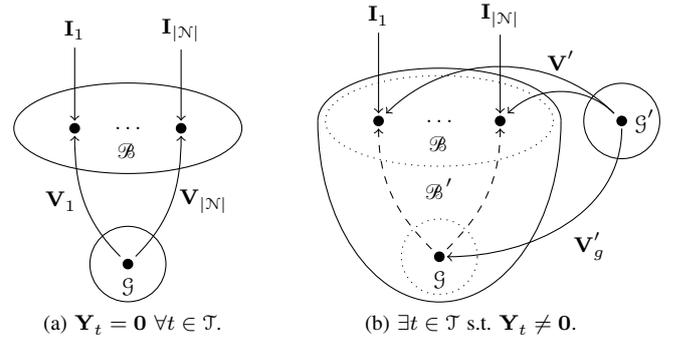


\begin{Theorem}
	\label{Thm:Rank}
	Suppose that Hypotheses \ref{Hyp:Neutral}--\ref{Hyp:Parameter} hold.
	If the branch graph $\Graph{B}=(\Set{N},\Set{L})$ is weakly connected, it follows that
	\begin{equation}
		\Rank{(\Tensor{Y})}
		=	\left\{
			\begin{array}{cc}
				(\Cardinality{\Set{N}}-1)\Cardinality{\Set{P}}
				&	\text{if}~\Tensor{Y}_{t}=\Tensor{0}~\forall t\in\Set{T}
				\\
				\Cardinality{\Set{N}}\Cardinality{\Set{P}}
				&	\text{otherwise}
			\end{array}
			\right.
		\label{Eq:Admittance:Rank}
	\end{equation}
	In other words, $\Tensor{Y}$ is non-singular if there is at least one non-zero shunt admittance in the circuit, and singular otherwise.
\end{Theorem}


The proof is conducted separately for the two cases.
\begin{Proof}[Case I: $\Tensor{Y}_{t}=\Tensor{0}~\forall t\in\Set{T}$, see Fig.~\ref{Fig:Grid:Original}.]
	From \eqref{Eq:Shunt:Admittance:Primitive}, it follows that $\Tensor{Y}_{\Set{T}}=\Tensor{0}$.
	Therefore, \eqref{Eq:Node:Admittance} simplifies to
	\begin{equation}
		\Tensor{Y}
		=	\Transpose{(\Tensor{A}_{\Graph{B}}^{\Set{P}})}
			\Tensor{Y}_{\Set{L}}
			\Tensor{A}_{\Graph{B}}^{\Set{P}}
	\end{equation}
	According to \eqref{Eq:Branch:Admittance:Primitive}, $\Tensor{Y}_{\Set{L}}$ is block diagonal.
	By Hypothesis~\ref{Hyp:Parameter}, its blocks $\Tensor{Y}_{\ell}$ ($\ell\in\Set{L}$) are symmetric and invertible.
	Therefore, $\Tensor{Y}_{\Set{L}}$ is also symmetric and invertible.
	According to Lemma~\ref{Lem:Factorization}, there exists a factorization of the form
	\begin{equation}
		\Tensor{Y}_{\Set{L}}
		=	\Transpose{\Tensor{U}_{\Set{L}}}\Tensor{D}_{\Set{L}}\Tensor{U}_{\Set{L}}
	\end{equation}
	where $\Tensor{U}_{\Set{L}}$ is unitary, and $\Tensor{D}_{\Set{L}}$ is positive diagonal.
	Therefore, there exists a matrix $\Tensor{E}_{\Set{L}}$ so that $\Tensor{D}_{\Set{L}}=\Transpose{\Tensor{E}_{\Set{L}}}\Tensor{E}_{\Set{L}}$, which is itself positive diagonal.
	Hence, $\Tensor{Y}$ can be written as
	\begin{equation}
		\Tensor{Y}
		=	\Transpose{\Tensor{M}_{\Set{L}}}\Tensor{M}_{\Set{L}},
		~	\Tensor{M}_{\Set{L}} = \Tensor{E}_{\Set{L}}\Tensor{U}_{\Set{L}}\Tensor{A}_{\Graph{B}}^{\Set{P}}
	\end{equation}
	Note that the term $\Tensor{E}_{\Set{L}}\Tensor{U}_{\Set{L}}$ is non-singular.
	Therefore, it follows from Lemmata~\ref{Lem:Rank:1} \& \ref{Lem:Rank:2} that
		\begin{equation}
			\Rank{(\Tensor{Y})}
			=	\Rank{(\Tensor{M}_{\Set{L}})}
			=	\Rank{(\Tensor{A}_{\Graph{B}}^{\Set{P}})}
		\end{equation}
	Recall the definition of $\Tensor{A}_{\Graph{B}}^{\Set{P}}$ \eqref{Eq:Incidence:Polyphase}.
	According to Lemma~\ref{Lem:Kronecker:Rank}
	\begin{align}
		\Rank(\Tensor{A}_{\Graph{B}}^{\Set{P}})
		&=	\Rank(\Tensor{A}_{\Graph{B}}) \cdot \Rank\left(\Diagonal(\Tensor{1}_{\Cardinality{\Set{P}}})\right)\\
		&=	\Rank(\Tensor{A}_{\Graph{B}}) \cdot \Cardinality{\Set{P}}
	\end{align}
	Since $\Graph{B}=(\Set{N},\Set{L})$ is weakly connected, $\Rank{(\Tensor{A}_{\Graph{B}})}=\Cardinality{\Set{N}}-1$ according to Lemma~\ref{Lem:Incidence:Rank}.
	This proves the claim.
	\qed
\end{Proof}


\begin{Proof}[Case II: $\exists t\in\Set{T}$ s.t. $\Tensor{Y}_{t}\neq\Tensor{0}$, see Fig.~\ref{Fig:Grid:Augmented}.]
	Introduce a virtual ground node $\Set{G}'$, and build a modified grid by turning the original ground node into another polyphase node.
	That is
	\begin{equation}
		\Set{N}' \coloneqq \Set{N} \cup \Set{G}
	\end{equation}
	Further, let $\Tensor{V}'$ be the counterpart of $\Tensor{V}$ referenced w.r.t. $\Set{G}'$, and $\Tensor{V}'_{g}$ the voltage of $\Set{G}$ w.r.t. $\Set{G}'$.
	Obviously, polyphase shunts with non-zero admittance in the original grid become polyphase branches in the modified grid.
	That is
	\begin{equation}
		\Set{L}' = \Set{L} \cup \{t\in\Set{T}~|~\Tensor{Y}_{t}\neq\Tensor{0}\}
	\end{equation}
	Hypothesis~\ref{Hyp:Parameter} implicates that the compound branch impedances $\Tensor{Z}_{\ell'}$ ($\ell'\in\Set{L}'$) are symmetric and invertible.
	By construction
	\begin{equation}
		\Tensor{Z}_{\ell'}
		=	\left\{
			\begin{array}{ll}
				\Tensor{Z}_{\ell}					&\text{if}~\ell'=\ell\in\Set{L}\\
				\Inverse{\Tensor{Y}_{t}}		&\text{if}~\ell'=t\in\Set{L}'\setminus\Set{L}
			\end{array}
			\right.
	\end{equation}
	The claimed properties follow from \eqref{Eq:Parameter:Branch} and \eqref{Eq:Parameter:Shunt}, respectively.
	Let $\Graph{B}'\coloneqq(\Set{N}',\Set{L}')$ be the analogon of $\Graph{B}$ for the modified grid.
	Obviously, if $\Graph{B}$ is weakly connected, $\Graph{B}'$ is weakly connected, too.
	Clearly, the modified grid satisfies the conditions required for the application of Theorem~\ref{Thm:Rank}.
	Observe that the compound shunt admittances from the polyphase nodes $\Set{N}'$ to the virtual ground node $\Set{G}'$ are zero by construction.
	\begin{equation}
		\Tensor{Y}_{t'} = \Tensor{0} ~ \forall t'\in\Set{T}'\coloneqq\Cartesian{\Set{N}'}{\Set{G}'}
	\end{equation}
	Therefore, the first part of \eqref{Eq:Admittance:Rank}, which has already been proven,
	can be applied.
	Accordingly, the compound nodal admittance matrix $\Tensor{Y}'$ of the modified grid has rank
	\begin{equation}
		\Rank{(\Tensor{Y}')}
		=	(\Cardinality{\Set{N}'}-1)\Cardinality{\Set{P}}
		=	\Cardinality{\Set{N}}\Cardinality{\Set{P}}
	\end{equation}
	Let $\Tensor{y}_{\Set{T}}$ be the column vector composed of the compound shunt admittances $\Tensor{Y}_{t}$ ($t\in\Set{T}$).
	Namely
	\begin{equation}
		\Tensor{y}_{\Set{T}} \coloneqq \Column_{t\in\Set{T}}(\Tensor{Y}_{t})
	\end{equation}
	Given that the voltages of the modified grid are referenced to the virtual ground node, Ohm's law \eqref{Eq:Ohm} reads as follows 
	\begin{equation}
		\renewcommand{\arraystretch}{1.5}
		\left[
		\begin{array}{l}
			\Tensor{I}\\
			\hline
			\Tensor{I}_{g}
		\end{array}
		\right]
		=	\left[
			\begin{array}{c|c}
				\Tensor{Y}										&-\Tensor{y}_{\Set{T}}\\
				\hline
				-\Transpose{\Tensor{y}_{\Set{T}}}		&\sum_{t\in\Set{T}}\Tensor{Y}_{t}
			\end{array}
			\right]
			\left[
			\begin{array}{l}
				\Tensor{V}'\\
				\hline
				\Tensor{V}'_{g}
			\end{array}
			\right]
	\end{equation}
	That is, $\Tensor{Y}'$ is given in block form as
	\begin{equation}
		\renewcommand{\arraystretch}{1.5}
		\Tensor{Y}'
		=	\left[
			\begin{array}{c|c}
				\Tensor{Y}										&-\Tensor{y}_{\Set{T}}\\
				\hline
				-\Transpose{\Tensor{y}_{\Set{T}}}		&\sum_{t\in\Set{T}}\Tensor{Y}_{t}
			\end{array}
			\right]
	\end{equation}
	It is known that elementary operations on the rows and column of a matrix do not change its rank.
	Hence, one can add the first $\Cardinality{\Set{N}}$ block rows/columns of $\Tensor{Y}'$ to the last block row/column without affecting its rank.
	Using Lemma~\ref{Lem:Admittance:Sum}, one finds
	\begin{align}
		\renewcommand{\arraystretch}{1.5}
		\left[
		\begin{array}{c|c}
			\Tensor{Y}										&-\Tensor{y}_{\Set{T}}\\
			\hline
			-\Transpose{\Tensor{y}_{\Set{T}}}		&\sum_{t\in\Set{T}}\Tensor{Y}_{t}
		\end{array}
		\right]
		&\left|~\Row_{\Cardinality{\Set{N}}+1} \pluseq \sum_{n\in\Set{N}}\Row_{n}\right.
		\\
		\left[
		\begin{array}{c|c}
			\Tensor{Y}		&-\Tensor{y}_{\Set{T}}\\
			\hline
			\Tensor{0}		&\Tensor{0}
		\end{array}
		\right]
		&\left|~\Column_{\Cardinality{\Set{N}}+1} \pluseq \sum_{n\in\Set{N}}\Column_{n}\right.
		\\
		\left[
		\begin{array}{c|c}
			\Tensor{Y}		&\Tensor{0}\\
			\hline
			\Tensor{0}		&\Tensor{0}
		\end{array}
		\right]
	\end{align}
	It follows straightforward that
	\begin{equation}
		\Rank(\Tensor{Y})
		=	\Rank(\Tensor{Y}')
		=	\Cardinality{\Set{N}}\Cardinality{\Set{P}}
	\end{equation}
	which proves the claim.
	\qed
\end{Proof}


\subsection{Block Rank}
\label{Sec:Properties:Blockrank}

\begin{Theorem}
	\label{Thm:Blockrank}
	Suppose that Hypotheses~\ref{Hyp:Neutral}--\ref{Hyp:Parameter} hold.
	If the branch graph $\Graph{B}=(\Set{N},\Set{L})$ is weakly connected, and all the compound branch impedances $\Tensor{Z}_{\ell}$ ($\ell\in\Set{L}$) are strictly passive
	\begin{equation}
		\Re\{\Tensor{Z}_{\ell}\}\succ0~\forall\ell\in\Set{L}
		\label{Eq:Blockrank:Condition}
	\end{equation}
	then it follows that
	\begin{equation}
		\Rank(\Tensor{Y}_{\Cartesian{\Set{M}}{\Set{M}}})
		=	\Cardinality{\Set{M}}\Cardinality{\Set{P}}
		~	\forall \Set{M}\subsetneq\Set{N}
		\label{Eq:Blockrank:Statement}
	\end{equation}
	That is, every proper diagonal subblock of the compound nodal admittance matrix $\Tensor{Y}$ has full rank.%
\end{Theorem}


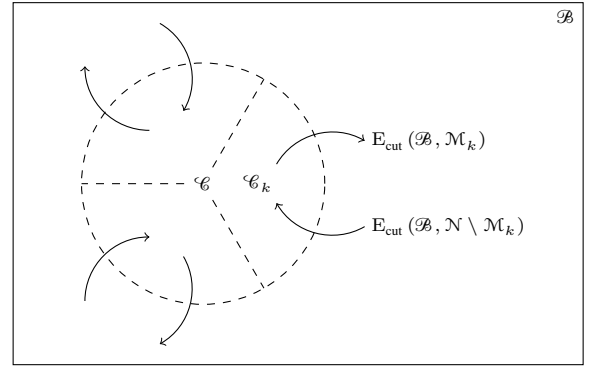
\begin{figure}[t]
	\centering
	\input{"Figures/Blockrank_Graph"}
	\caption{Branch graphs and cut-sets in the fictional grid used for the proof of Theorem~\ref{Thm:Blockrank} (surfaces indicate weakly connected graphs).}
	\label{Fig:Blockrank:Graph}
\end{figure}

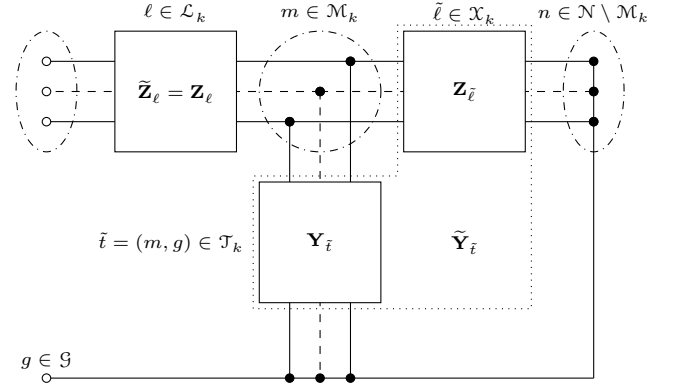
\begin{figure}[t]
	\centering
	\input{"Figures/Blockrank_Circuit"}	
	\caption{Compound branch impedances and shunt admittances in the fictional grid used for the proof of Theorem~\ref{Thm:Blockrank}.}
	\label{Fig:Blockrank:Circuit}
\end{figure}


\begin{Proof}
	As known from circuit theory, $\Tensor{Y}_{\Cartesian{\Set{M}}{\Set{M}}}$ relates $\Tensor{I}_{\Set{M}}$ and $\Tensor{V}_{\Set{M}}$ for $\Tensor{V}_{\Set{N}\setminus\Set{M}}=\Tensor{0}$ (i.e., the polyphase nodes $\Set{N}\setminus\Set{M}$ are short-circuited).
	This is due to the assumption that the state of each electrical component is composed solely of ground-referenced nodal voltages.
	In that sense, $\Tensor{Y}_{\Cartesian{\Set{M}}{\Set{M}}}$ can be regarded as the compound nodal admittance matrix of a fictional grid which is obtained by grounding the polyphase nodes $\Set{N}\setminus\Set{M}$.
	Therefore, the polyphase branches internal to $\Set{M}$ (i.e., $\Internal{\Graph{B}}{\Set{M}}$) persist, whereas those interconnecting $\Set{M}$ and $\Set{N}\setminus\Set{M}$ (i.e., $\Cut{\Graph{B}}{\Set{M}}$ and $\Cut{\Graph{B}}{\Set{N}\setminus\Set{M}}$) become polyphase shunts in the modified grid.
	The remaining branches are described by the graph
	\begin{equation}
		\Graph{C} \coloneqq (\Set{M},\Internal{\Graph{B}}{\Set{M}})
	\end{equation}
	In general, $\Graph{C}$ is disconnected.
	However, since $\Graph{B}$ is weakly connected as a whole, there exists a partition $\{\Set{M}_{k}\,|\,k\in\Set{K}\}$ of $\Set{M}$ so that the subgraphs $\Graph{C}_{k}$ associated with $\Set{M}_{k}$
	\begin{equation}
		\Graph{C}_{k}\coloneqq(\Set{M}_{k},\Internal{\Graph{B}}{\Set{M}_{k}}),~k\in\Set{K}
	\end{equation}		
	are weakly connected, and mutually disconnected (see Fig.~\ref{Fig:Blockrank:Graph}).
	Therefore, $\Tensor{Y}_{\Cartesian{\Set{M}}{\Set{M}}}$ is block diagonal.
	Namely
	\begin{equation}
		\Tensor{Y}_{\Cartesian{\Set{M}}{\Set{M}}}
		=	\Diagonal_{k\in\Set{K}}(\Tensor{Y}_{\Cartesian{\Set{M}_{k}}{\Set{M}_{k}}})
	\end{equation}
	In turn, $\Tensor{Y}_{\Cartesian{\Set{M}_{k}}{\Set{M}_{k}}}$ can be interpreted as the compound nodal admittance matrix of a fictional grid, which is constructed by grounding the polyphase nodes $\Set{N}\setminus\Set{M}_{k}$.
	Define
	\begin{align}
		\Set{L}_{k}
		&\coloneqq	\Edges{\Graph{C}_{k}} = \Internal{\Graph{B}}{\Set{M}_{k}} \subset \Set{L}\\
		\Set{T}_{k}	
		&\coloneqq	\Cartesian{\Set{M}_{k}}{\Set{G}} \subset \Set{T}
	\end{align}
	
	
	The compound branch impedances $\widetilde{\Tensor{Z}}_{\ell}$ and compound shunt admittances $\widetilde{\Tensor{Y}}_{t}$ of this fictional grid satisfy Hypothesis~\ref{Hyp:Parameter}.
	The grounding process does not affect the compound branch impedances.
	Therefore (see Fig.~\ref{Fig:Blockrank:Circuit})
	\begin{align}
		\ell\in\Set{L}_{k}:~\widetilde{\Tensor{Z}}_{\ell}=\Tensor{Z}_{\ell}
	\end{align}
	Therefore, \eqref{Eq:Parameter:Branch} of Hypothesis~\ref{Hyp:Parameter} obviously applies to $\widetilde{\Tensor{Z}}_{\ell}$, too.
	In contrast, the compound shunt admittances are modified, as some of the polyphase branches become polyphase shunts.
	The set $\Set{X}_{k}$ containing these polyphase branches is given by
	\begin{equation}
		\Set{X}_{k} \coloneqq \Cut{\Graph{B}}{\Set{M}_{k}} \cup \Cut{\Graph{B}}{\Set{N}\setminus\Set{M}_{k}}
	\end{equation}
	Since $\Graph{B}$ is weakly connected, it holds that $\Set{X}_{k}\neq\emptyset$~$\forall k\in\Set{K}$.
	Therefore, $\exists\tilde{\ell}\in\Set{X}_{k}$ $\forall k\in\Set{K}$, which is a polyphase branch in the original grid, and a polyphase shunt in the modified grid.
	The compound branch impedances of the grounded polyphase branches contribute to the compound branch admittances of the modified grid.
	Namely (see Fig.~\ref{Fig:Blockrank:Circuit})
	\begin{equation}
		\left[~
		\begin{aligned}
			t&\in\Set{T}_{k},\\
			t&=(m,g)
		\end{aligned}
		\right.
		:~
		\widetilde{\Tensor{Y}}_{t}
		=		\Tensor{Y}_{t}
				+	\sum\limits_{\substack{\ell\in\Set{X}_{k}\\\ell=(m,n)}}\Tensor{Y}_{\ell}
				+	\sum\limits_{\substack{\ell\in\Set{X}_{k}\\\ell=(n,m)}}\Tensor{Y}_{\ell}
	\end{equation}
	If the sums are empty, \eqref{Eq:Parameter:Shunt} of Hypothesis~\ref{Hyp:Parameter} obviously applies.
	Otherwise, it follows from Hypothesis~\ref{Hyp:Parameter}, \eqref{Eq:Blockrank:Condition}, and Lemma~\ref{Lem:Definite:Real} that $\widetilde{\Tensor{Y}}_{t}$ is symmetric and invertible, and has positive definite real part.
	Hence, \eqref{Eq:Parameter:Shunt} of Hypothesis~\ref{Hyp:Parameter} applies in this case, too.
	
	
	As Hypothesis~\ref{Hyp:Parameter} holds, Theorem~\ref{Thm:Rank} can be applied.
	Due to the fact that $\Set{X}_{k}\neq\emptyset$ $\forall k\in\Set{K}$, it follows that $\exists \tilde{t}\in\Set{T}_{k}$ $\forall k\in\Set{K}$, for which $\widetilde{\Tensor{Y}}_{\tilde{t}}\neq\Tensor{0}$, even if $\Tensor{Y}_{t}=\Tensor{0}$ $\forall t\in\Set{T}_{k}$ (see Fig.~\ref{Fig:Blockrank:Circuit}).
	Thus
	\begin{equation}
		\Rank(\Tensor{Y}_{\Cartesian{\Set{M}_{k}}{\Set{M}_{k}}})
		= 	\Cardinality{\Set{M}_{k}}\Cardinality{\Set{P}}
		~	\forall k\in\Set{K}
	\end{equation}
	Since $\Tensor{Y}_{\Cartesian{\Set{M}}{\Set{M}}}$ is block diagonal with blocks $\Tensor{Y}_{\Set{M}_{k}\times\Set{M}_{k}}$
	\begin{align}
		\Rank(\Tensor{Y}_{\Cartesian{\Set{M}}{\Set{M}}})
		&=	\sum\limits_{k\in\Set{K}}\Rank(\Tensor{Y}_{\Cartesian{\Set{M}_{k}}{\Set{M}_{k}}})\\
		&=	\Cardinality{\Set{P}}\sum\limits_{k\in\Set{K}}\Cardinality{\Set{M}_{k}}
		=		\Cardinality{\Set{P}}\Cardinality{\Set{M}}
	\end{align}
	This proves the claim.
	\qed
\end{Proof}


%% file: Figures/Rank_Graph.tex
\begin{tikzpicture}
	\footnotesize
		
	
	\def\BusPosition{0.7}
	\def\BusSpacing{0.3}
	\def\GroundHeight{-1.8}
	\def\CurrentHeight{+1.3}
	
	
	\node at (-\BusPosition,0) (BusLeft) {};
	\node at (\BusPosition,0) (BusRight) {};
	\node at (0,0) (BusCentre) {$\ldots$};
	\node at (0,\GroundHeight) (Ground) {};
	
	\fill (BusLeft) circle [radius=2pt];
	\fill (BusRight) circle [radius=2pt];
	\fill (Ground) circle [radius=2pt];
	
	\draw (Ground) [->,out=135,in=-90] to node[midway,left]{$\Tensor{V}_{1}$} (BusLeft);
	\draw (Ground) [->,out=45,in=-90] to node[midway,right]{$\Tensor{V}_{\Cardinality{\Set{N}}}$} (BusRight);
	
	
	\node at (-\BusPosition,\CurrentHeight) (CurrentLeft) {$\Tensor{I}_{1}$};
	\node at (\BusPosition,\CurrentHeight) (CurrentRight) {$\Tensor{I}_{\Cardinality{\Set{N}}}$};
	
	\draw (CurrentLeft) [->] to (BusLeft);
	\draw (CurrentRight) [->] to (BusRight);
	
	
	\draw (BusCentre) ellipse (1.5 and 0.6);
	\draw (Ground) circle (0.5);
	
	\node at ($(BusCentre)-(0,0.3)$) {$\Graph{B}$};
	\node at ($(Ground)-(0,0.3)$) {$\Set{G}$};
	
\end{tikzpicture}

%% file: Figures/Rank_Graph_Augmented.tex
\tikzstyle{block} = [draw, rectangle, minimum width = 0.75cm, minimum height = 0.75cm]
\tikzstyle{sum} = [draw, circle, minimum size=.5cm, node distance=1.75cm]
\tikzstyle{input} = [coordinate]
\tikzstyle{output} = [coordinate]

\tikzset{
    partial ellipse/.style args={#1:#2:#3}{
        insert path={+ (#1:#3) arc (#1:#2:#3)}
    }
}

\begin{tikzpicture}
	\footnotesize
		
	
	\def\BusPosition{+0.8}
	\def\GroundHeight{-1.8}
	\def\CurrentHeight{+1.4}
	\def\VirtualGroundPosition{+2.4}
	
	
	\node at (-\BusPosition,0) (BusLeft) {};
	\node at (\BusPosition,0) (BusRight) {};
	\node at (0,0) (BusCentre){$\ldots$};
	\node at (0,\GroundHeight) (GroundPhysical) {};
	\node at (\VirtualGroundPosition,0) (GroundVirtual) {};
	
	\fill (BusLeft) circle [radius=2pt];
	\fill (BusRight) circle [radius=2pt];
	\fill (GroundPhysical) circle [radius=2pt];
	\fill (GroundVirtual) circle [radius=2pt];
	
	\draw[dashed] (GroundPhysical) [->,out=135,in=-90] to node[midway,left]{} (BusLeft);
	\draw[dashed] (GroundPhysical) [->,out=45,in=-90] to node[midway,right]{} (BusRight);
	
	
	\draw (GroundVirtual) [->,out=135,in=45] to node [near start,above]{$\Tensor{V}'$} (BusLeft);
	\draw (GroundVirtual) [->,out=135,in=45] to (BusRight);
	\draw (GroundVirtual) [->,out=-90,in=0] to node[midway,below right]{$\Tensor{V}'_{g}$}  (GroundPhysical);
	
	
	\node at (-\BusPosition,\CurrentHeight) (CurrentLeft) {$\Tensor{I}_{1}$};
	\node at (\BusPosition,\CurrentHeight) (CurrentRight) {$\Tensor{I}_{\Cardinality{\Set{N}}}$};
	
	\draw (CurrentLeft) [->] to (BusLeft);
	\draw (CurrentRight) [->] to (BusRight);
	
	
	\draw[dotted] (BusCentre) ellipse (1.5 and 0.6);
	\draw[dotted] (GroundPhysical) circle (0.5);
	\draw (GroundVirtual) circle (0.5);
	\draw (BusCentre) [partial ellipse=0:180:1.6 and 0.7];
	\draw (BusCentre) [partial ellipse=180:360:1.6 and 2.4];
	
	\node at ($(0,\GroundHeight/2)$){$\Graph{B}'$};
	\node at ($(GroundVirtual)+(0.3,0)$) {$\Set{G}'$};
	
	\node at ($(BusCentre)-(0,0.3)$) {$\Graph{B}$};
	\node at ($(Ground)-(0,0.3)$) {$\Set{G}$};
	
\end{tikzpicture}

%% file: Figures/Blockrank_Graph.tex
\begin{tikzpicture}
	\scriptsize
		
	\def\xLeft{-2.5}
	\def\xRight{5.0}
	\def\y{2.4}
	\def\Radius{1.6}
	\def\RadiusInner{1.0}
	\def\RadiusOuter{2.2}

	
	\coordinate (LU) at (\xLeft,\y);	
	\coordinate (LL) at (\xLeft,-\y);
	\coordinate (RU) at (\xRight,\y);	
	\coordinate (RL) at (\xRight,-\y);
	
	\draw (LU) to (LL) to (RL) to (RU) to cycle;
	
	\node[below left] at (RU) {$\Graph{B}$};
	
	\draw[dashed] (0,0) circle (\Radius);
	\foreach \phi in {60,180,300} {
		\draw[dashed] ($(\phi:0.25)$) -- ($(\phi:\Radius)$);
	}
	
	\node at (0,0) {$\Graph{C}$};
	\node[left] at ($(0:\RadiusInner)$) {$\Graph{C}_{k}$};
	
	\node[right] at ($(15:\RadiusOuter)$) {$\Cut{\Graph{B}}{\Set{M}_{k}}$};
	\node[right] at ($(345:\RadiusOuter)$) {$\Cut{\Graph{B}}{\Set{N}\setminus\Set{M}_{k}}$};
	
	\foreach \phi in {15,135,255} {
		\draw ($(\phi:\RadiusInner)$) [->,out=\phi+45,in=180+\phi-45] to ($(\phi:\RadiusOuter)$);
	}
	
	\foreach \phi in {105,225,345} {
		\draw ($(\phi:\RadiusInner)$) [<-,out=\phi-45,in=180+\phi+45] to ($(\phi:\RadiusOuter)$);
	}
\end{tikzpicture}

%% file: Figures/Blockrank_Circuit.tex
\begin{circuitikz}
	\scriptsize	
		
	\def\BranchPosition{1.9} 
	\def\BranchHeight{3.8} 
	\def\ShuntHeight{1.8} 
	\def\PortPosition{3.6} 
	\def\BlockSize{1.6} 
	\def\WireSpacing{0.4} 
	
	\coordinate (PhaseLeft) at (-\PortPosition,\BranchHeight);	
	\coordinate (PhaseRight) at (\PortPosition,\BranchHeight);
	
	
	\coordinate (NeutralLeft) at (-\PortPosition,0);	
	\coordinate (NeutralRight) at (\PortPosition,0);
	
	\draw (NeutralLeft) to[short,o-] (NeutralRight);
	\node at ($(NeutralLeft)+0.6*(0,\WireSpacing)$) {$g\in\Set{G}$};
	
	
	\coordinate (ShuntPhase) at (0,\BranchHeight);
	\coordinate (ShuntNeutral) at (0,0);
	
	\draw[dashed] (ShuntPhase) to (ShuntNeutral);	
	\draw (ShuntPhase) to[open,*-*] (ShuntNeutral);
	\draw ($(ShuntPhase)+\WireSpacing*(1,1)$) to[short,*-*] ($(ShuntNeutral)+\WireSpacing*(1,0)$);
	\draw ($(ShuntPhase)-\WireSpacing*(1,1)$) to[short,*-*] ($(ShuntNeutral)-\WireSpacing*(1,0)$);
	
	\coordinate (Shunt) at (0,\ShuntHeight);
	\draw[fill=white] ($(Shunt)-0.5*\BlockSize*(1,1)$) rectangle ($(Shunt)+0.5*\BlockSize*(1,1)$);
	\node at (Shunt) {$\Tensor{Y}_{\tilde{t}}$};
	\node at ($(Shunt)-4.9*(\WireSpacing,0)$) {$\tilde{t}=(m,g)\in\Set{T}_{k}$};
	
	\draw[dashdotted] (ShuntPhase) circle (2*\WireSpacing);
	\node at ($(ShuntPhase)+2.6*(0,\WireSpacing)$) {$m\in\Set{M}_{k}$};
	
	
	\draw[dashed] (ShuntPhase) to (PhaseLeft);
	\draw (ShuntPhase) to[open,*-o] (PhaseLeft);
	\draw($(ShuntPhase)+\WireSpacing*(1,1)$) to[short,*-o] ($(PhaseLeft)+\WireSpacing*(0,1)$);	
	\draw($(ShuntPhase)-\WireSpacing*(1,1)$) to[short,*-o] ($(PhaseLeft)-\WireSpacing*(0,1)$);
	
	\coordinate (BranchLeft) at (-\BranchPosition,\BranchHeight);
	\draw[fill=white] ($(BranchLeft)-0.5*\BlockSize*(1,1)$) rectangle ($(BranchLeft)+0.5*\BlockSize*(1,1)$);
	\node at (BranchLeft) {$\widetilde{\Tensor{Z}}_{\ell}=\Tensor{Z}_{\ell}$};
	\node at ($(BranchLeft)+2.6*(0,\WireSpacing)$) {$\ell\in\Set{L}_{k}$};
	
	\draw[dashdotted] (PhaseLeft) ellipse (0.4 and 0.8);
	
	
	\draw[dashed] (ShuntPhase) to[short] (PhaseRight);
	\draw ($(ShuntPhase)+\WireSpacing*(1,1)$) to[short] ($(PhaseRight)+\WireSpacing*(0,1)$);
	\draw ($(ShuntPhase)-\WireSpacing*(1,1)$) to[short] ($(PhaseRight)-\WireSpacing*(0,1)$);
	\draw ($(PhaseRight)+\WireSpacing*(0,1)$) to[short,*-*] (PhaseRight) to[short,*-*] ($(PhaseRight)-\WireSpacing*(0,1)$) to[short,*-] (NeutralRight);	
	
	\coordinate (BranchRight) at (\BranchPosition,\BranchHeight);
	\draw[fill=white] ($(BranchRight)-0.5*\BlockSize*(1,1)$) rectangle ($(BranchRight)+0.5*\BlockSize*(1,1)$);
	\node at (BranchRight) {$\Tensor{Z}_{\tilde{\ell}}$};
	\node at ($(BranchRight)+2.6*(0,\WireSpacing)$) {$\tilde{\ell}\in\Set{X}_{k}$};
	
	\draw[dashdotted] (PhaseRight) ellipse (0.4 and 0.8);	
	\node at ($(PhaseRight)+2.6*(0,\WireSpacing)$) {$n\in\Set{N}\setminus\Set{M}_{k}$};
	
	\draw[dotted] ($(BranchRight)+2.2*\WireSpacing*(1,1)$)
		to ($(\BranchPosition,\ShuntHeight)+2.2*\WireSpacing*(1,-1)$)
		to ($(Shunt)-2.2*\WireSpacing*(1,1)$)
		to ($(Shunt)+2.2*\WireSpacing*(-1,1)$)
		to (\BranchPosition-2.2*\WireSpacing,\ShuntHeight+2.2*\WireSpacing)
		to ($(BranchRight)+2.2*\WireSpacing*(-1,1)$)
		to cycle;
	\node at (\BranchPosition,\ShuntHeight) {$\widetilde{\Tensor{Y}}_{\tilde{t}}$};
	
\end{circuitikz}

%% file: Sections/Implications.tex
\section{Implications}
\label{Sec:Implications}

Using the properties developed in the pevious section, the findings of \cite{J:CT:YM:2017:Kettner} can be extended to polyphase power systems.
In Sec.~\ref{Sec:Implications:Kron}, it is is proven that Kron reduction is feasible for any subset of the polyphase nodes with zero current injections.
In Sec.~\ref{Sec:Implications:Hybrid}, it is shown that hybrid parameters matrices exist for arbitrary partitions of the polyphase nodes.


\WhiteSpaceMagic{-0.3cm}

\subsection{Kron Reduction}
\label{Sec:Implications:Kron}

Ohm's law \eqref{Eq:Ohm} establishes the link between injected current phasors and nodal voltage phasors through the grid.
Obviously, the current injected into a polyphase node also depends on the devices which are connected there.
In particular, a polyphase node $z\in\Set{N}$, in which no devices are present, has zero injected current (i.e., $\Tensor{I}_{z}=\Tensor{0}$). 
Let $\Set{Z}\subsetneq\Set{N}$ ($\Set{Z}\neq\emptyset$) be a set of such \emph{zero-injection nodes} (i.e., $\Tensor{I}_{\Set{Z}}=\Tensor{0}$).
As known from power system analysis, \eqref{Eq:Ohm} can be reduced by eliminating the zero-injection nodes $\Set{Z}$ through \emph{Kron reduction} (see \cite{B:CT:1959:Kron}).
This yields a model of the grid with fewer unknowns.
Therefore, computationally heavy tasks like \PFS, \SE, and \VSA can be accelerated without using high-performance computers (see \cite{J:PFS:1992:Ajjarapu,J:PSSE:1971:Mendel,J:PSSE:2017:Kettner,J:PSSA:1993:Loef,J:PSSA:1997:Irisarri,B:HPC:2013:Khaitan}).


In order for Kron reduction to be applicable, the diagonal subblock $\Tensor{Y}_{\Cartesian{\Set{Z}}{\Set{Z}}}$ of $\Tensor{Y}$ has to be invertible (this will be shown shortly).
Interestingly, researchers and practitioners hardly ever check whether this condition is actually satisfied.
In fact, even the inventor (i.e., \cite{B:CT:1959:Kron}) did not consider this issue.
According to experience, Kron reduction is feasible in practice, but there was no theoretical proof for this empirical observation until recently.
Lately, some researchers (i.e., \cite{J:CT:KR:2013:Doerfler,J:CT:YM:2017:Kettner}) examined this problem for monophase grids.
The results of these works do apply to balanced triphase grids (respectively, the equivalent positive-sequence networks), but not to generic unbalanced polyphase grids.
This gap in the existing literature is closed by Corollary~\ref{Cor:Kron}, which provides a guarantee for the feasibility of Kron reduction in generic polyphase grids.


\begin{Corollary}
	\label{Cor:Kron}
	Suppose that Theorem~\ref{Thm:Blockrank} applies.
	Let $\Set{Z}\subsetneq\Set{N}$ be a non-empty subset of $\Set{N}$ with zero current injection.
	\begin{equation}
		\Tensor{I}_{\Set{Z}} = \Tensor{0}
		\label{Eq:Kron:Condition}
	\end{equation}
	Then, the voltages at the zero-injection buses $\Set{Z}$ linearly depend on those at the other buses $\Complement{\Set{Z}}\coloneqq\Set{N}\setminus\Set{Z}$, and \eqref{Eq:Ohm} reduces to
	\begin{equation}
		\Tensor{I}_{\Complement{\Set{Z}}} = \widehat{\Tensor{Y}}\Tensor{V}_{\Complement{\Set{Z}}}
		,~
		\widehat{\Tensor{Y}}
		\coloneqq	\Schur{\Tensor{Y}}{\Tensor{Y}_{\Cartesian{\Set{Z}}{\Set{Z}}}}
		\label{Eq:Kron:Admittance}
	\end{equation}
	The reduced compound nodal admittance matrix $\widehat{\Tensor{Y}}$ satisfies
	\begin{equation}
		\Rank(\widehat{\Tensor{Y}}_{\Cartesian{\Set{M}}{\Set{M}}})
		=	\Cardinality{\Set{M}}\Cardinality{\Set{P}}
		~	\forall\Set{M}\subsetneq\Complement{\Set{Z}}
		\label{Eq:Kron:Rank}
	\end{equation}
	That is, $\widehat{\Tensor{Y}}$ has the same properties as $\Tensor{Y}$ w.r.t. the rank of its proper diagonal subblocks.
\end{Corollary}


Observe that, according to \eqref{Eq:Kron:Condition}, a zero-injection node has zero current injection in every phase.
Moreover, according to \eqref{Eq:Kron:Admittance}, $\widehat{\Tensor{Y}}$ is the Schur complement of $\Tensor{Y}$ w.r.t. $\Tensor{Y}_{\Cartesian{\Set{Z}}{\Set{Z}}}$.
In order for this Schur complement to exist, $\Tensor{Y}_{\Cartesian{\Set{Z}}{\Set{Z}}}$ has to be invertible.


\begin{Proof}
	Observe that $\Set{Z}$ and $\Complement{\Set{Z}}$ form a partition of $\Set{N}$.
	Therefore, Ohm's law \eqref{Eq:Ohm} can be written in block form as
	\begin{equation}
		\left[
		\begin{array}{l}
			\Tensor{I}_{\Complement{\Set{Z}}}\\
			\Tensor{I}_{\Set{Z}}
		\end{array}
		\right]
		=	\left[
			\begin{array}{ll}
					\Tensor{Y}_{\Cartesian{\Complement{\Set{Z}}}{\Complement{\Set{Z}}}}
				&	\Tensor{Y}_{\Cartesian{\Complement{\Set{Z}}}{\Set{Z}}}
				\\
					\Tensor{Y}_{\Cartesian{\Set{Z}}{\Complement{\Set{Z}}}}
				&	\Tensor{Y}_{\Cartesian{\Set{Z}}{\Set{Z}}}
			\end{array}
			\right]
			\left[
			\begin{array}{l}
				\Tensor{V}_{\Set{Z}_{\complement}}\\
				\Tensor{V}_{\Set{Z}}
			\end{array}
			\right]
	\end{equation}
	According to Theorem~\ref{Thm:Blockrank}, $\Tensor{Y}_{\Cartesian{\Set{Z}}{\Set{Z}}}$ has full rank.
	Therefore
	\begin{equation}
		\exists\Tensor{Z}_{\Cartesian{\Set{Z}}{\Set{Z}}}
		\coloneqq	\Inverse{\Tensor{Y}_{\Cartesian{\Set{Z}}{\Set{Z}}}}
	\end{equation}
	From $\Tensor{I}_{\Set{Z}}=\Tensor{0}$, it follows straightforward that
	\begin{equation}
		\Tensor{V}_{\Set{Z}}
		=	-\Inverse{\Tensor{Y}}_{\Cartesian{\Set{Z}}{\Set{Z}}}
			\Tensor{Y}_{\Cartesian{\Set{Z}}{\Complement{\Set{Z}}}}
			\Tensor{V}_{\Complement{\Set{Z}}}
	\end{equation}
	So, $\Tensor{V}_{\Set{Z}}$ is a linear function of $\Tensor{V}_{\Complement{\Set{Z}}}$ as claimed.
	Substitute this formula back into Ohm's law, and obtain
	\begin{align}
		\Tensor{I}_{\Complement{\Set{Z}}}
		&=		\Tensor{Y}_{\Cartesian{\Complement{\Set{Z}}}{\Complement{\Set{Z}}}}\Tensor{V}_{\Set{Z}_{\complement}}
				+	\Tensor{Y}_{\Cartesian{\Complement{\Set{Z}}}{\Set{Z}}}\Tensor{V}_{\Set{Z}}
		\\
		&=
					\Tensor{Y}_{\Complement{\Set{Z}}\times\Complement{\Set{Z}}}
					\Tensor{V}_{\Set{Z}_{\complement}}
				-	\Tensor{Y}_{\Cartesian{\Complement{\Set{Z}}}{\Set{Z}}}
					\Inverse{\Tensor{Y}}_{\Cartesian{\Set{Z}}{\Set{Z}}}
					\Tensor{Y}_{\Cartesian{\Set{Z}}{\Complement{\Set{Z}}}}
					\Tensor{V}_{\Complement{\Set{Z}}}
		\\
		&=	(\Schur{\Tensor{Y}}{\Tensor{Y}_{\Cartesian{\Set{Z}}{\Set{Z}}}})\Tensor{V}_{\Complement{\Set{Z}}}
	\end{align}
	This proves the first claim \eqref{Eq:Kron:Admittance}.
	According to Lemma~\ref{Lem:Schur:Blockform}, the diagonal block of $\widehat{\Tensor{Y}}$ associated with $\Set{M}\subsetneq\Complement{\Set{Z}}$ is given by
	\begin{align}
		\widehat{\Tensor{Y}}_{\Cartesian{\Set{M}}{\Set{M}}}
		&=	(\Schur{\Tensor{Y}}{\Tensor{Y}_{\Cartesian{\Set{Z}}{\Set{Z}}}})_{\Cartesian{\Set{M}}{\Set{M}}}
		\\
		&=	\left[
				\begin{array}{ll}
						\Tensor{Y}_{\Cartesian{\Set{M}}{\Set{M}}}
					&	\Tensor{Y}_{\Cartesian{\Set{M}}{\Set{Z}}}
					\\
						\Tensor{Y}_{\Cartesian{\Set{Z}}{\Set{M}}}
					&	\Tensor{Y}_{\Cartesian{\Set{Z}}{\Set{Z}}}
				\end{array}
				\right]
				/~\Tensor{Y}_{\Cartesian{\Set{Z}}{\Set{Z}}}
		\\
		&=		\Tensor{Y}_{\Cartesian{(\Set{M}\cup\Set{Z})}{(\Set{M}\cup\Set{Z})}}
				/	\Tensor{Y}_{\Cartesian{\Set{Z}}{\Set{Z}}}
	\end{align}
	From $\Set{M}\subsetneq\Complement{\Set{Z}}=\Set{N}\setminus\Set{Z}$ and $\Set{Z}\subsetneq\Set{N}$, it follows that $\Set{M}\cup\Set{Z}\subsetneq\Set{N}$.
	That is, both $\Set{M}\cup\Set{Z}$ and $\Set{Z}$ are proper subsets of $\Set{N}$.
	According to Theorem~\ref{Thm:Blockrank}, the diagonal subblocks of $\Tensor{Y}$ associated with $\Set{M}\cup\Set{Z}$ and $\Set{Z}$ therefore have full rank.
	By consequence
	\begin{align}
		D_{\Set{M}\cup\Set{Z}}
		&\coloneqq	\det(\Tensor{Y}_{\Cartesian{(\Set{M}\cup\Set{Z})}{(\Set{M}\cup\Set{Z})}}) \neq 0
		\\
		D_{\Set{Z}}
		&\coloneqq	\det(\Tensor{Y}_{\Cartesian{\Set{Z}}{\Set{Z}}}) \neq 0
	\end{align}
	From Lemma~\ref{Lem:Schur:Determinant}, it follows that
	\begin{equation}
		\det(\widehat{\Tensor{Y}}_{\Cartesian{\Set{M}}{\Set{M}}})
		=		D_{\Set{M}\cup\Set{Z}} \cdot D_{\Set{Z}} \neq 0
	\end{equation}
	Therefore, $\widehat{\Tensor{Y}}_{\Cartesian{\Set{M}}{\Set{M}}}$ has full rank.
	Namely
	\begin{equation}
		\Rank(\widehat{\Tensor{Y}}_{\Cartesian{\Set{M}}{\Set{M}}})
		=	\Cardinality{\Set{M}}\Cardinality{\Set{P}}
	\end{equation}
	As $\Set{M}\subsetneq\Complement{\Set{Z}}$ is arbitrary, this proves the second claim \eqref{Eq:Kron:Rank}.
	\qed
\end{Proof}


Moreover, Corollary~\ref{Cor:Kron} itself has a fundamental implication.
\begin{Observation}
	\label{Obs:Kron}
	Kron reduction preserves the property which enables its applicability in the first place (i.e., that every proper diagonal subblock of a compound nodal admittance matrix has full rank).
	Therefore, if $\Set{Z}$ is partitioned as $\{\Set{Z}_{k} \mid k\in\Set{K}\}$, the parts $\Set{Z}_{k}$ can be reduced one after another, and the (partially or fully) reduced compound nodal admittance matrices obtained after every step of the reduction also exhibit the rank property.
\end{Observation}
\noindent
Performing the reduction sequentially rather than ``en bloc'' is beneficial in terms of computational burden, because the Schur complement \eqref{Eq:Schur} requires a matrix inversion.
This operation is computationally expensive, and scales poorly with problem size (even if the inverse is not computed explicitly).


\WhiteSpaceMagic{-0.3cm}

\subsection{Hybrid Parameters}
\label{Sec:Implications:Hybrid}

Evidently, the circuit equations \eqref{Eq:Ohm} are in \emph{admittance form}.
Namely, the injected current and nodal voltage phasors appear in separate vectors, which are linked by the nodal admittance matrix $\Tensor{Y}$.
In power system analysis, it is often more convenient to write the circuit equations in \emph{hybrid form} (if this is feasible).
The corresponding system of linear equations features vectors composed of both voltage and current phasors, and is described by a so-called \emph{hybrid parameters matrix} $\Tensor{H}$.
In this context, observe that there is no guarantee for the existence of a hybrid representation.
Whether a suitable matrix $\Tensor{H}$ exists, depends both on the nodal admittance matrix $\Tensor{Y}$ of the grid and on the partition of the nodes underlying the hybrid representation.


Various researchers have treated the subject of hybrid parameters matrices for monophase grids.
Some authors plainly describe how a hybrid parameters matrix can be built, provided that it exists at all (e.g., \cite{J:CT:HM:1977:Haji,J:CT:HM:1990:Sun,J:CT:HM:1991:Augusto}).
Others do provide criteria for the existence of hybrid parameters matrices, but only for some (i.e., at least one) partition of the nodes (e.g., \cite{J:CT:HM:1965:So,J:CT:HM:1965:Zuidweg,J:CT:HM:1966:Anderson}).
One recent work establishes a criterion for arbitrary partitions of the nodes \cite{J:CT:YM:2017:Kettner}.
All of the aforementioned works only study the monophase case.
Accordingly, those results may apply to balanced triphase grids (respectively, their equivalent positive-sequence networks), but not to generic unbalanced polyphase grids.
In contrast, Corollary~\ref{Cor:Hybrid} ensures the existence of hybrid parameters matrices for arbitrary partitions of the polyphase nodes of a generic polyphase grid.


\begin{Corollary}
	\label{Cor:Hybrid}
	Suppose that Theorem \ref{Thm:Blockrank} applies.
	Let $\Set{M}\subsetneq\Set{N}$ be non-empty, that is $\Set{M}$ and $\Complement{\Set{M}}\coloneqq\Set{N}\setminus\Set{M}$ form a partition of $\Set{N}$.
	Then, there exists a \emph{compound hybrid parameters matrix $\Tensor{H}$}, and the grid is described by the hybrid multiport equations
	\begin{equation}
		\left[~
		\begin{aligned}
			\Tensor{I}	&{}_{\Complement{\Set{M}}}\\
			\Tensor{V}	&{}_{\Set{M}}
		\end{aligned}
		~\right]
		=	\Tensor{H}
			\left[~
			\begin{aligned}
				\Tensor{V}	&{}_{\Complement{\Set{M}}}\\
				\Tensor{I}	&{}_{\Set{M}}
			\end{aligned}
			~\right]
	\end{equation}
	The blocks of $\Tensor{H}$ are given as follows
	\begin{align}
		\Tensor{H}_{\Cartesian{\Set{M}}{\Set{M}}}
		&=	\Inverse{\Tensor{Y}_{\Cartesian{\Set{M}}{\Set{M}}}}
		\label{Eq:Hybrid:1}
		\\
		\Tensor{H}_{\Cartesian{\Set{M}}{\Complement{\Set{M}}}}
		&=	-\Inverse{\Tensor{Y}_{\Cartesian{\Set{M}}{\Set{M}}}}
				\Tensor{Y}_{\Cartesian{\Set{M}}{\Complement{\Set{M}}}}
		\label{Eq:Hybrid:2}
		\\
		\Tensor{H}_{\Cartesian{\Complement{\Set{M}}}{\Set{M}}}
		&=	\Tensor{Y}_{\Cartesian{\Complement{\Set{M}}}{\Set{M}}}
				\Inverse{\Tensor{Y}_{\Cartesian{\Set{M}}{\Set{M}}}}
		\label{Eq:Hybrid:3}
		\\
		\Tensor{H}_{\Cartesian{\Complement{\Set{M}}}{\Complement{\Set{M}}}}
		&=	\Schur{\Tensor{Y}}{\Tensor{Y}_{\Cartesian{\Set{M}}{\Set{M}}}}
		\label{Eq:Hybrid:4}
	\end{align}
\end{Corollary}

\begin{Proof}
	Write Ohm's law \eqref{Eq:Ohm} for $\Tensor{I}_{\Set{M}}$
	\begin{equation}
		\Tensor{I}_{\Set{M}}
		=		\Tensor{Y}_{\Cartesian{\Set{M}}{\Complement{\Set{M}}}}\Tensor{V}_{\Complement{\Set{M}}}
			+	\Tensor{Y}_{\Cartesian{\Set{M}}{\Set{M}}}\Tensor{V}_{\Set{M}}
	\end{equation}
	By Theorem~\ref{Thm:Blockrank}, $\Tensor{Y}_{\Cartesian{\Set{M}}{\Set{M}}}$ has full rank, and is hence invertible.
	Define $\Tensor{Z}_{\Cartesian{\Set{M}}{\Set{M}}} \coloneqq \Inverse{\Tensor{Y}_{\Cartesian{\Set{M}}{\Set{M}}}}$, and solve for $\Tensor{V}_{\Set{M}}$.
	\begin{align}
		\Tensor{V}_{\Set{M}}
		&=	\Inverse{\Tensor{Y}}_{\Cartesian{\Set{M}}{\Set{M}}}
				(
						\Tensor{I}_{\Set{M}}
					-	\Tensor{Y}_{\Cartesian{\Set{M}}{\Complement{\Set{M}}}}\Tensor{V}_{\Complement{\Set{M}}}
				)
		\\
		&=		\Tensor{H}_{\Cartesian{\Set{M}}{\Set{M}}}\Tensor{I}_{\Set{M}}
				+	\Tensor{H}_{\Cartesian{\Set{M}}{\Complement{\Set{M}}}}\Tensor{V}_{\Complement{\Set{M}}}
	\end{align}
	as claimed in	 \eqref{Eq:Hybrid:1}--\eqref{Eq:Hybrid:2}.
	Write Ohm's law \eqref{Eq:Ohm} for $\Tensor{I}_{\Complement{\Set{M}}}$, and substitute the above formula for $\Tensor{V}_{\Set{M}}$.
	This gives 
	\begin{align}
		\Tensor{I}_{\Complement{\Set{M}}}
		&=		\Tensor{Y}_{\Cartesian{\Complement{\Set{M}}}{\Complement{\Set{M}}}}\Tensor{V}_{\Complement{\Set{M}}}
				-	\Tensor{Y}_{\Cartesian{\Complement{\Set{M}}}{\Set{M}}}\Tensor{V}_{\Set{M}}
		\\
		&=	\left[~
				\begin{aligned}
						&	\Tensor{Y}_{\Cartesian{\Complement{\Set{M}}}{\Set{M}}}
							\Inverse{\Tensor{Y}}_{\Cartesian{\Set{M}}{\Set{M}}}
							\Tensor{I}_{\Set{M}}
					\\
					+	&	\Tensor{Y}_{\Cartesian{\Complement{\Set{M}}}{\Complement{\Set{M}}}}
							\Tensor{V}_{\Complement{\Set{M}}}
					\\
					-	&	\Tensor{Y}_{\Cartesian{\Complement{\Set{M}}}{\Set{M}}}
							\Inverse{\Tensor{Y}}_{\Cartesian{\Set{M}}{\Set{M}}}
							\Tensor{Y}_{\Cartesian{\Set{M}}{\Complement{\Set{M}}}}
							\Tensor{V}_{\Complement{\Set{M}}}
				\end{aligned}
				\right.
		\\
		&=		\Tensor{H}_{\Cartesian{\Complement{\Set{M}}}{\Set{M}}}
					\Tensor{I}_{\Set{M}}
				+	\Tensor{H}_{\Cartesian{\Complement{\Set{M}}}{\Complement{\Set{M}}}}
					\Tensor{V}_{\Complement{\Set{M}}}
	\end{align}
	as claimed in \eqref{Eq:Hybrid:3}--\eqref{Eq:Hybrid:4}.
	\qed
\end{Proof}

It is worth noting that compound hybrid parameters matrices also exist for Kron-reduced grids.
\begin{Observation}
	\label{Obs:Hybrid}
	The existence of compound hybrid parameters matrices is based on the same rank property as the feasibility of Kron reduction.
	Since Kron reduction preserves this property (recall Observation~\ref{Obs:Kron}), compound hybrid parameters matrices can be obtained from unreduced, partially reduced, and fully reduced compound nodal admittance matrices.
	In this regard, $\Tensor{Y}$ can be replaced by $\widehat{\Tensor{Y}}$ in Corollary~\ref{Cor:Hybrid}.
\end{Observation}

%% file: Sections/Conclusions.tex
\section{Conclusions}
\label{Sec:Conclusions}

This paper examined the properties of the compound nodal admittance matrix of polyphase power systems, and illustrated their implications for power system analysis.
Using the concept of compound electrical parameters, and exploiting the physical characteristics of electrical components, rank properties for the compound nodal admittance matrix and its diagonal subblocks were deduced.
Notably, it was proven that the diagonal blocks have full rank if the grid is connected and lossy.
Based on these findings, it was shown that the feasibility of Kron reduction and the existence of hybrid parameters are guaranteed in practice.
Thus, this paper provided a rigorous theoretical foundation for the analysis of generic polyphase power systems.

%% file: Sections/Equipment.tex
\section{Power System Components}
\label{App:Equipment}

	
\subsection{Transmission Lines}
\label{App:Model:Line}

Consider a transmission line with $\Cardinality{\Set{P}}$ phase conductors and one neutral conductor.
Let $\Tensor{v}(z,t)$ and $\Tensor{i}(z,t)$ be the vectors of phase-to-neutral voltages and phase conductor currents, i.e.
\begin{align}
	\Tensor{v}(z,t)
	&=	\Column_{p\in\Set{P}}\left(v_{p}(z,t)\right)
	\\
	\Tensor{i}(z,t)
	&=	\Column_{p\in\Set{P}}\left(i_{p}(z,t)\right)
\end{align}
where $z$  is the position along the line.
If (i) the electromagnetic parameters of the line are state-independent, (ii) the conductors are parallel, and the perpendicular distance between any two of them is much shorter than the wavelength, (iii) the conductors have finite conductance, (iv) the electromagnetic field outside of the conductors produced by the charges and currents inside of them is purely transversal, and (v) the sum of the conductor currents is zero, Maxwell's equations simplify to the so-called \emph{telegrapher's equations} (see \cite{B:PSA:2008:Paul})
\begin{align}
	\partial_{z}\Tensor{v}(z,t)
	&=	-(\Tensor{R}'+\Tensor{L}'\partial_{t})\Tensor{i}(z,t)
	\\
	\partial_{z}\Tensor{i}(z,t)
	&=	-(\Tensor{G}'+\Tensor{C}'\partial_{t})\Tensor{v}(z,t)
\end{align}
$\Tensor{R}'$ and $\Tensor{L}'$ are the resistance and inductance per unit length of the conductors,
$\Tensor{G}'$ and $\Tensor{C}'$ are the conductance and capacitance per unit length of the dielectric.
These matrices are symmetric.
Consider a segment of infinitesimal length $\Delta z$ at position $z$.
Let $E_{e}(z,t)$ and $E_{m}(z,t)$ be the energy stored in the \emph{electric} and \emph{magnetic field}, respectively.
They are given by (see \cite{B:PSA:2008:Paul})
\begin{align}
	E_{e}(z,t)
	&=	\frac{\Delta z}{2}\Transpose{(\Tensor{v}(z,t))}\Tensor{C}'\Tensor{v}(z,t)
	\\
	E_{m}(z,t)
	&=	\frac{\Delta z}{2}\Transpose{(\Tensor{i}(z,t))}\Tensor{L}'\Tensor{i}(z,t)
\end{align}
As $E_{e}(z,t)>0~\forall\Tensor{v}(z,t)\neq\Tensor{0}$ and $E_{m}(z,t)>0~\forall\Tensor{i}(z,t)\neq\Tensor{0}$, it follows straightforward that $\Tensor{C}'\succ0$ and $\Tensor{L}'\succ0$.
Similarly, let $P_{c}(z,t)$ and $P_{d}(z,t)$ be the power dissipated in the conductors and the ambient \emph{dielectric}.
They are given by (see \cite{B:PSA:2008:Paul})
\begin{align}
	P_{d}(z,t)
	&=	\Delta z \cdot \Transpose{(\Tensor{v}(z,t))}\Tensor{G}'\Tensor{v}(z,t)
	\\
	P_{c}(z,t)
	&=	\Delta z \cdot \Transpose{(\Tensor{i}(z,t))}\Tensor{R}'\Tensor{i}(z,t)
\end{align}
Since real systems are lossy, $\Tensor{P}_{d}(z,t)>0~\forall\Tensor{v}(z,t)\neq\Tensor{0}$ and $\Tensor{P}_{c}(z,t)>0~\forall\Tensor{i}(z,t)\neq\Tensor{0}$.
Therefore, $\Tensor{G}'\succ0$ and $\Tensor{R}'\succ0$.
If a transmission line is \emph{electrically short} (i.e., its total length $|z_{m}-z_{n}|$ is significantly shorter than the wavelength), it can be represented by a $\Pie$-section equivalent circuit with
\begin{align}
	\Tensor{Y}_{\Pie,m|(m,n)}
	&=	\frac{1}{2}(\Tensor{G}'+j\omega\Tensor{C}')|z_{m}-z_{n}|
	\\
	\Tensor{Y}_{\Pie,n|(m,n)}
	&=	\Tensor{Y}_{\Pie,m|(m,n)}
	\\
	\Tensor{Z}_{\Pie,(m,n)}
	&=	(\Tensor{R}'+j\omega\Tensor{L}')|z_{m}-z_{n}|
\end{align}
Since $\Tensor{G}'$, $\Tensor{C}'$, $\Tensor{R}'$, and $\Tensor{L}'$ are real positive definite, $\Tensor{Y}_{\Pie,m|(m,n)}$, $\Tensor{Y}_{\Pie,n|(m,n)}$, and $\Tensor{Z}_{\Pie,(m,n)}$ are symmetric with positive definite real part.
So, according to Lemmata~\ref{Lem:Definite:Real} \& \ref{Lem:Definite:Imaginary}, they are invertible.
This is in accordance with Hypothesis~\ref{Hyp:Parameter}.


\subsection{Transformers}
\label{App:Model:Transformer}

When analyzing power systems operating at rated frequency, transformers are represented by a $\Tea$-section equivalent circuits, which correspond one-to-one to the composition of the devices (see \cite{B:PSA:2011:Heathcote}).
Consider a polyphase transformer connecting two polyphase nodes $m,n\in\Set{N}$.
Let $m$ be its \emph{primary side}, and $n$ its \emph{secondary side}.
In this case, the parameters of the $\Tea$-section equivalent circuit are given as
\begin{align}
	\Tensor{Z}_{\Tea,(m,x)}
	&=	\Tensor{R}_{w,1} + j\omega\Tensor{L}_{\ell,1}
	\\
	\Tensor{Z}_{\Tea,(x,n)}
	&=	\Tensor{R}_{w,2} + j\omega\Tensor{L}_{\ell,2}
	\\
	\Tensor{Y}_{\Tea,x}
	&=	\Tensor{G}_{h} + j\omega\Tensor{B}_{m}
\end{align}
The resistance matrices $\Tensor{R}_{w,1}$ and $\Tensor{R}_{w,2}$ represent the \emph{winding resistances}, and the inductance matrices $\Tensor{L}_{\ell,1}$ and $\Tensor{L}_{\ell,2}$ the \emph{leakage inductances} of the coils on the primary and secondary side, respectively.
The former are positive diagonal, the latter are positive definite.
The conductance matrix $\Tensor{G}_{h}$ represents the \emph{hysteresis losses}, and the susceptance matrix $\Tensor{B}_{m}$ the \emph{magnetization} of the transformer's core.
The former is positive diagonal, the latter is positive definite.
Therefore, according to Lemmata~\ref{Lem:Definite:Real} \& \ref{Lem:Definite:Imaginary}, $\Tensor{Z}_{\Tea,(m,x)}$, $\Tensor{Z}_{\Tea,(x,n)}$, and $\Tensor{Y}_{\Tea,x}$ are invertible.
This holds irrespective of whether the polyphase transformer is built with one single multi-leg core or several separate cores.
In the latter case, $\Tensor{L}_{\ell,1}$, $\Tensor{L}_{\ell,2}$, and $\Tensor{B}_{m}$ are diagonal, since separate cores are magnetically decoupled.
In either case, the compound electrical parameters satisfy Hypothesis~\ref{Hyp:Parameter}.


\subsection{\FACTS Devices}
\label{App:Model:FACTS}

\begin{figure}[t]
	\centering
	\subfloat[Series compensation.]
	{%
		\centering
		\input{"Figures/FACTS_Series_Compensator"}
		\label{Fig:FACTS:Series}
	}
	
	\subfloat[Shunt compensation (for simplicity, only one compensator is shown).]
	{%
		\centering
		\input{"Figures/FACTS_Shunt_Compensator"}
		\label{Fig:FACTS:Shunt}
	}	
	
	\caption{Incorporation of \FACTS devices into the transmission line model.}
	\label{Fig:FACTS}
\end{figure}
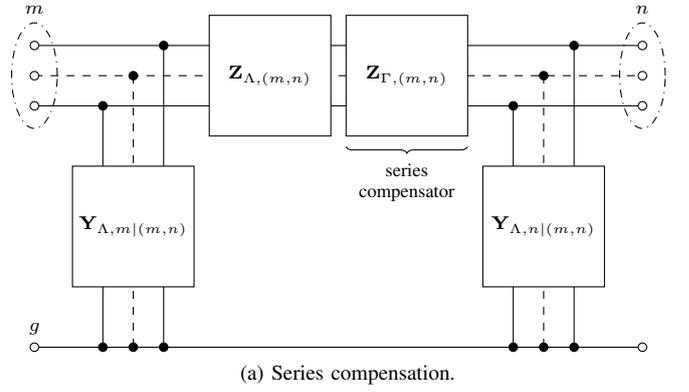
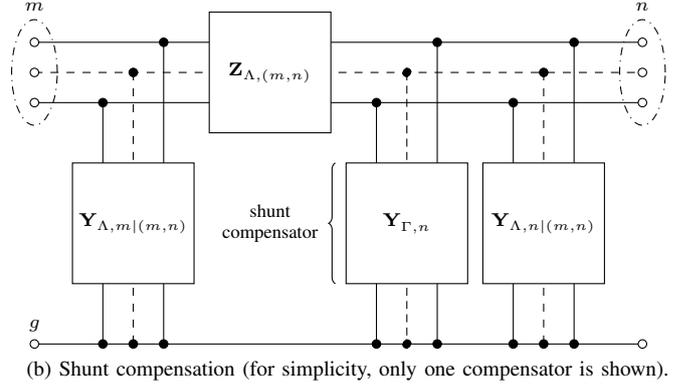

There exist three families of FACTS devices, namely \emph{series compensators}, \emph{shunt compensators}, and combined \emph{series-and-shunt compensators} (see \cite{J:PSA:2001:Gerbex}).
Here, the first two cases are discussed.
Let $\Tensor{Z}_{\Lambda,(m,n)}$, $\Tensor{Y}_{\Lambda,m|(m,n)}$, and $\Tensor{Y}_{\Lambda,n|(m,n)}$ be the compound electrical parameters describing a transmission line without compensation (see App.~\ref{App:Model:Line}).
If a series compensator is installed, the compound branch impedance of the respective transmission line is altered.
In the $\Pie$-section equivalent circuit, this is reflected by adding the compound impedance $\Tensor{Z}_{\Gamma,(m,n)}$  of the compensator to the compound branch impedance of the transmission line (see Fig.~\ref{Fig:FACTS:Series}).
Namely
\begin{equation}
	\Tensor{Z}_{\Pie,(m,n)} = \Tensor{Z}_{\Lambda,(m,n)} + \Tensor{Z}_{\Gamma,(m,n)}
	\label{Eq:FACTS:Zmn}
\end{equation}
If shunt compensators are installed, the compound admittances $\Tensor{Y}_{\Gamma,m}$ and $\Tensor{Y}_{\Gamma,n}$ of the compensators add to the compound shunt admittance of the transmission line  (see Fig.~\ref{Fig:FACTS:Shunt}).
Namely
\begin{align}
	\Tensor{Y}_{\Pie,m|(m,n)}
	&=	\Tensor{Y}_{\Lambda,m|(m,n)} + \Tensor{Y}_{\Gamma,m}
	\label{Eq:FACTS:Ym}
	\\
	\Tensor{Y}_{\Pie,n|(m,n)}
	&=	\Tensor{Y}_{\Lambda,n|(m,n)} + \Tensor{Y}_{\Gamma,n}
	\label{Eq:FACTS:Yn}
\end{align}%
Usually, such compensators are built from banks of capacitors or inductors, which can be stepwise (dis)connected.
This kind of devices are symmetrical w.r.t. the phases, and lossy.
Hence
\begin{align}
	\text{series compensation}
	&:	\left[~
		\begin{aligned}
			\Tensor{Z}_{\Gamma,(m,n)}	 &= \Transpose{\Tensor{Z}_{\Gamma,(m,n)}}\\
			\Re\{\Tensor{Z}_{\Gamma,(m,n)}\}&\succ0
		\end{aligned}
		\right.
	\\
	\text{shunt compensation}
	&:	\left[~
		\begin{aligned}
			\Tensor{Y}_{\Gamma,m/n}	 &= \Transpose{\Tensor{Y}_{\Gamma,m/n}}\\
			\Re\{\Tensor{Y}_{\Gamma,m/n}\}&\succ0
		\end{aligned}
		\right.
\end{align}
Since the compound electrical parameters of transmission lines satisfy Hypothesis~\ref{Hyp:Parameter} (see App.~\ref{App:Model:Line}), the compound electrical parameters \eqref{Eq:FACTS:Zmn} and \eqref{Eq:FACTS:Ym}--\eqref{Eq:FACTS:Yn} are symmetric, have positive definite real part, and are invertible (by Lemma~\ref{Lem:Definite:Real}).
Accordingly, transmission lines equipped with series or shunt compensators satisfy Hypothesis~\ref{Hyp:Parameter}.

%% file: Figures/FACTS_Series_Compensator.tex
\begin{circuitikz}
	\scriptsize
	
	
	\def\PortPosition{4.0}
	\def\PortHeight{3.6}
	\def\BranchPosition{0.9}	
	\def\ShuntPosition{2.7}
	\def\ShuntHeight{1.6}
	\def\BlockSize{1.6}
	\def\WireSpacing{0.4}
		
	
	
	\coordinate (PhaseLeft) at (-\PortPosition,\PortHeight);
	\coordinate (PhaseRight) at (\PortPosition,\PortHeight);
	\coordinate (NeutralLeft) at (-\PortPosition,0);
	\coordinate (NeutralRight) at (\PortPosition,0);
	
	\draw[dashed] (PhaseLeft) to[short] (PhaseRight);
	\draw (PhaseLeft) to[open,o-o] (PhaseRight);
	\draw ($(PhaseLeft)+\WireSpacing*(0,1)$) to[short,o-o] ($(PhaseRight)+\WireSpacing*(0,1)$);
	\draw ($(PhaseLeft)-\WireSpacing*(0,1)$) to[short,o-o] ($(PhaseRight)-\WireSpacing*(0,1)$);
	\draw (NeutralLeft) to[short,o-o] (NeutralRight);
	
	\coordinate (Branch) at (-\BranchPosition,\PortHeight);
	\draw[fill=white] ($(Branch)-0.5*\BlockSize*(1,1)$) rectangle ($(Branch)+0.5*\BlockSize*(1,1)$);
	\node at (Branch) {$\Tensor{Z}_{\Lambda,(m,n)}$};
	
	\coordinate (Compensator) at (\BranchPosition,\PortHeight);
	\draw[fill=white] ($(Compensator)-0.5*\BlockSize*(1,1)$) rectangle ($(Compensator)+0.5*\BlockSize*(1,1)$);
	\node at (Compensator) {$\Tensor{Z}_{\Gamma,(m,n)}$};
	
	\draw[dashdotted] (PhaseLeft) ellipse (0.3 and 0.7) node at ($(PhaseLeft)+2.2*(0,\WireSpacing)$) {$m$};
	\draw[dashdotted] (PhaseRight) ellipse (0.3 and 0.7) node at ($(PhaseRight)+2.2*(0,\WireSpacing)$) {$n$};
	\node at ($(NeutralLeft)+0.6*(0,\WireSpacing)$) {$g$};
	\draw[decoration={brace,mirror,raise=4pt},decorate] ($(Compensator)-0.5*\BlockSize*(1,1)$) to node[below=7pt] {\begin{tabular}{c}series\\compensator\end{tabular}} ($(Compensator)+0.5*\BlockSize*(1,-1)$);
	
	
	
	\coordinate (ShuntLeftPhase) at (-\ShuntPosition,\PortHeight);
	\coordinate (ShuntLeftNeutral) at (-\ShuntPosition,0);
	
	\draw[dashed] (ShuntLeftPhase) to[short] (ShuntLeftNeutral);
	\draw (ShuntLeftPhase) to[open,*-*] (ShuntLeftNeutral);
	\draw ($(ShuntLeftPhase)+\WireSpacing*(1,1)$) to[short,*-*] ($(ShuntLeftNeutral)+\WireSpacing*(1,0)$);
	\draw ($(ShuntLeftPhase)-\WireSpacing*(1,1)$) to[short,*-*] ($(ShuntLeftNeutral)-\WireSpacing*(1,0)$);
	
	\coordinate (ShuntLeft) at (-\ShuntPosition,\ShuntHeight);
	\draw[fill=white] ($(ShuntLeft)-0.5*\BlockSize*(1,1)$) rectangle ($(ShuntLeft)+0.5*\BlockSize*(1,1)$);
	\node at (ShuntLeft) {$\Tensor{Y}_{\Lambda,m|(m,n)}$};
		
	\coordinate (ShuntRightPhase) at (\ShuntPosition,\PortHeight);
	\coordinate (ShuntRightNeutral) at (\ShuntPosition,0);
	
	\draw[dashed] (ShuntRightPhase) to[short] (ShuntRightNeutral);
	\draw (ShuntRightPhase) to[open,*-*] (ShuntRightNeutral);
	\draw ($(ShuntRightPhase)+\WireSpacing*(1,1)$) to[short,*-*] ($(ShuntRightNeutral)+\WireSpacing*(1,0)$);
	\draw ($(ShuntRightPhase)-\WireSpacing*(1,1)$) to[short,*-*] ($(ShuntRightNeutral)-\WireSpacing*(1,0)$);
	
	\coordinate (ShuntRight) at (\ShuntPosition,\ShuntHeight);
	\draw[fill=white] ($(ShuntRight)-0.5*\BlockSize*(1,1)$) rectangle ($(ShuntRight)+0.5*\BlockSize*(1,1)$);
	\node at (ShuntRight) {$\Tensor{Y}_{\Lambda,n|(m,n)}$};
	
\end{circuitikz}

%% file: Figures/FACTS_Shunt_Compensator.tex
\begin{circuitikz}
	\scriptsize	
		
	
	\def\PortPosition{4.0}
	\def\PortHeight{3.6}
	\def\BranchPosition{0.9}	
	\def\ShuntPosition{2.7}
	\def\ShuntHeight{1.6}
	\def\BlockSize{1.6}
	\def\WireSpacing{0.4}
	
	
	
	\coordinate (PhaseLeft) at (-\PortPosition,\PortHeight);
	\coordinate (PhaseRight) at (\PortPosition,\PortHeight);
	\coordinate (NeutralLeft) at (-\PortPosition,0);
	\coordinate (NeutralRight) at (\PortPosition,0);
	
	\draw[dashed] (PhaseLeft) to[short] (PhaseRight);
	\draw (PhaseLeft) to[open,o-o] (PhaseRight);
	\draw ($(PhaseLeft)+\WireSpacing*(0,1)$) to[short,o-o] ($(PhaseRight)+\WireSpacing*(0,1)$);
	\draw ($(PhaseLeft)-\WireSpacing*(0,1)$) to[short,o-o] ($(PhaseRight)-\WireSpacing*(0,1)$);
	\draw (NeutralLeft) to[short,o-o] (NeutralRight);
	
	\coordinate (Branch) at (-\BranchPosition,\PortHeight);
	\draw[fill=white] ($(Branch)-0.5*\BlockSize*(1,1)$) rectangle ($(Branch)+0.5*\BlockSize*(1,1)$);
	\node at (Branch) {$\Tensor{Z}_{\Lambda,(m,n)}$};
	
	\draw[dashdotted] (PhaseLeft) ellipse (0.3 and 0.7) node at ($(PhaseLeft)+2.2*(0,\WireSpacing)$) {$m$};
	\draw[dashdotted] (PhaseRight) ellipse (0.3 and 0.7) node at ($(PhaseRight)+2.2*(0,\WireSpacing)$) {$n$};
	\node at ($(NeutralLeft)+0.6*(0,\WireSpacing)$) {$g$};
	
	
	
	\coordinate (CompensatorPhase) at (\BranchPosition,\PortHeight);
	\coordinate (CompensatorNeutral) at (\BranchPosition,0);
	
	\draw[densely dashed] (CompensatorPhase) to[short] (CompensatorNeutral);
	\draw[densely dashed] (CompensatorPhase) to[open,*-*] (CompensatorNeutral);
	\draw ($(CompensatorPhase)+\WireSpacing*(1,1)$) to[short,*-*] ($(CompensatorNeutral)+\WireSpacing*(1,0)$);
	\draw ($(CompensatorPhase)-\WireSpacing*(1,1)$) to[short,*-*] ($(CompensatorNeutral)-\WireSpacing*(1,0)$);
	
	\coordinate (Compensator) at (\BranchPosition,\ShuntHeight);
	\draw[fill=white] ($(Compensator)-0.5*\BlockSize*(1,1)$) rectangle ($(Compensator)+0.5*\BlockSize*(1,1)$);
	\node at (Compensator) {$\Tensor{Y}_{\Gamma,n}$};
	
	\draw[decoration={brace,mirror,raise=4pt},decorate] ($(Compensator)+0.5*\BlockSize*(-1,1)$) to node[left=2pt] {\begin{tabular}{c}shunt\\compensator\end{tabular}} ($(Compensator)+0.5*\BlockSize*(-1,-1)$);
	
	
	
	\coordinate (ShuntLeftPhase) at (-\ShuntPosition,\PortHeight);
	\coordinate (ShuntLeftNeutral) at (-\ShuntPosition,0);
	
	\draw[dashed] (ShuntLeftPhase) to[short] (ShuntLeftNeutral);
	\draw (ShuntLeftPhase) to[open,*-*] (ShuntLeftNeutral);
	\draw ($(ShuntLeftPhase)+\WireSpacing*(1,1)$) to[short,*-*] ($(ShuntLeftNeutral)+\WireSpacing*(1,0)$);
	\draw ($(ShuntLeftPhase)-\WireSpacing*(1,1)$) to[short,*-*] ($(ShuntLeftNeutral)-\WireSpacing*(1,0)$);
	
	\coordinate (ShuntLeft) at (-\ShuntPosition,\ShuntHeight);
	\draw[fill=white] ($(ShuntLeft)-0.5*\BlockSize*(1,1)$) rectangle ($(ShuntLeft)+0.5*\BlockSize*(1,1)$);
	\node at (ShuntLeft) {$\Tensor{Y}_{\Lambda,m|(m,n)}$};
		
	\coordinate (ShuntRightPhase) at (\ShuntPosition,\PortHeight);
	\coordinate (ShuntRightNeutral) at (\ShuntPosition,0);
	
	\draw[dashed] (ShuntRightPhase) to[short] (ShuntRightNeutral);
	\draw (ShuntRightPhase) to[open,*-*] (ShuntRightNeutral);
	\draw ($(ShuntRightPhase)+\WireSpacing*(1,1)$) to[short,*-*] ($(ShuntRightNeutral)+\WireSpacing*(1,0)$);
	\draw ($(ShuntRightPhase)-\WireSpacing*(1,1)$) to[short,*-*] ($(ShuntRightNeutral)-\WireSpacing*(1,0)$);
	
	\coordinate (ShuntRight) at (\ShuntPosition,\ShuntHeight);
	\draw[fill=white] ($(ShuntRight)-0.5*\BlockSize*(1,1)$) rectangle ($(ShuntRight)+0.5*\BlockSize*(1,1)$);
	\node at (ShuntRight) {$\Tensor{Y}_{\Lambda,n|(m,n)}$};
		
\end{circuitikz}

%% file: Sections/Biographies.tex

\begin{IEEEbiography}[{\includegraphics[width=1in,height=1.25in,,keepaspectratio]{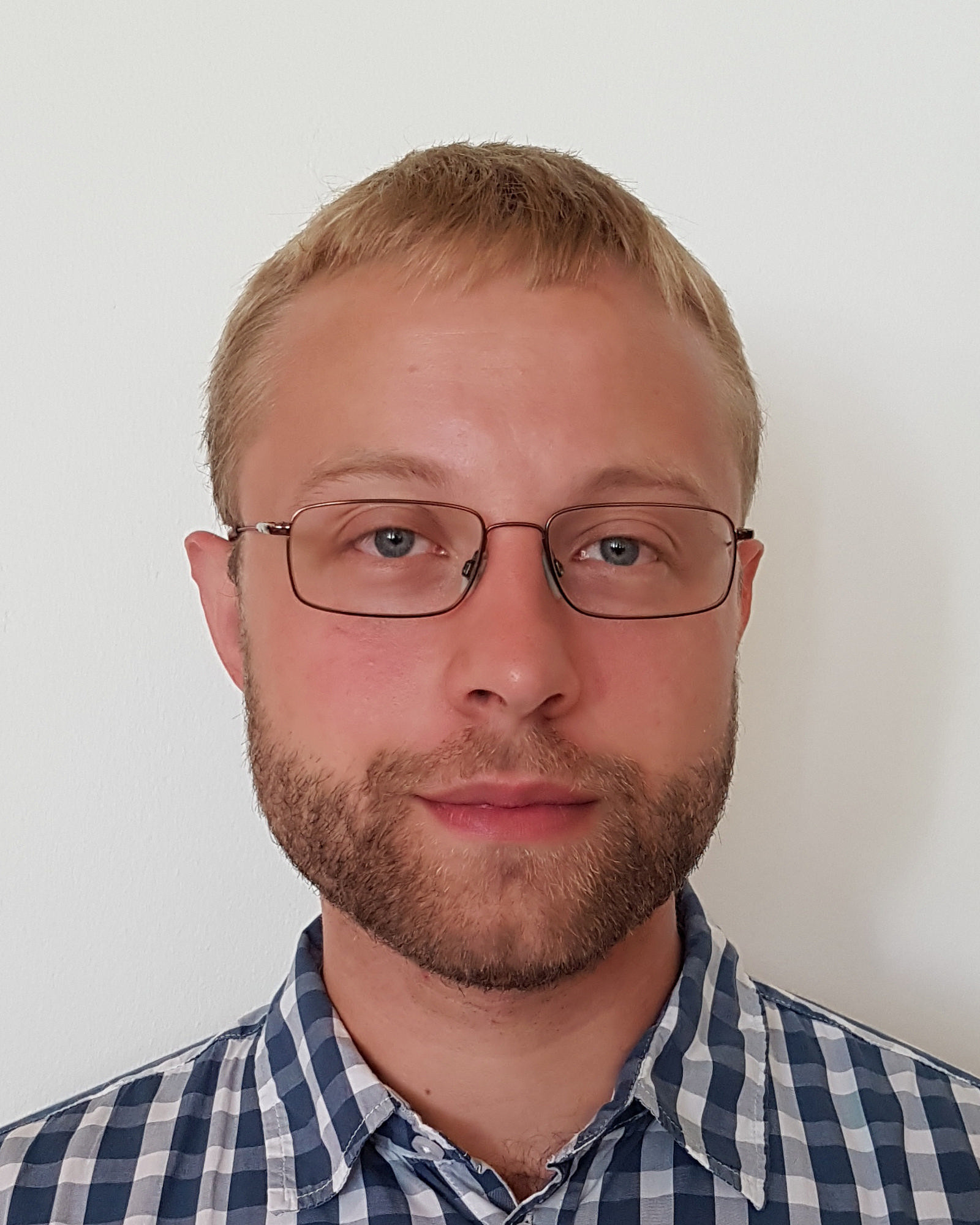}}]{Andreas Martin Kettner}
	(M'15) received the B.Sc. and M.Sc. degrees in electrical engineering and information technology from the Swiss Federal Institute of Technology of Z{\"u}rich (ETHZ), Z{\"u}rich, Switzerland, in 2012 and 2014, respectively.
	After working as a Development Engineer for Supercomputing Systems AG in Z{\"u}rich, he joined the Distributed Electrical Systems Laboratory at the Swiss Federal Institute of Technology of Lausanne (EPFL), Lausanne, Switzerland, where he is pursuing a Ph.D. degree.
	His research is focused on real-time situation awareness for monitoring and control of active distribution networks, with special reference to power system state estimation and voltage stability assessment.
\end{IEEEbiography}



\begin{IEEEbiography}[{\includegraphics[width=1in,height=1.25in,keepaspectratio]{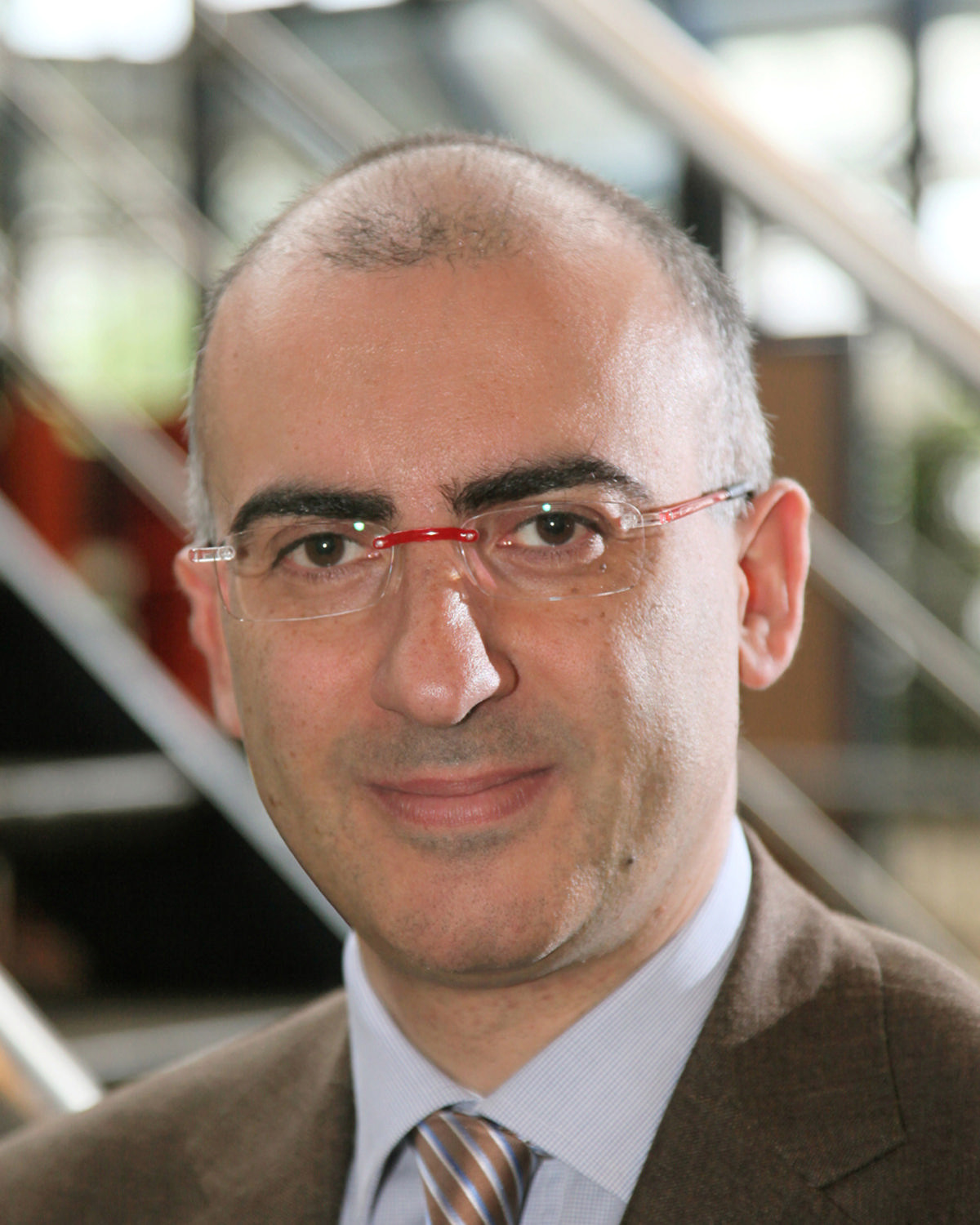}}]{Mario Paolone}
	(M'07-SM'10) received the M.Sc. (with Hons.) and Ph.D. degrees in electrical engineering from the University of Bologna, Italy, in 1998 and 2002, respectively.
	In 2005, he was appointed Assistant Professor in power systems with the University of Bologna, where he was with the Power Systems Laboratory until 2011.
	In 2010, he received the Associate Professor eligibility from the Polytechnic of Milan, Italy.
	Since 2011, he joined the Swiss Federal Institute of Technology of Lausanne (EPFL), Lausanne, Switzerland, where he is currently Full Professor, Chair of the Distributed Electrical Systems Laboratory (DESL), Head of SCCER--FURIES (Swiss Competence Center for Energy Research, Future Swiss Electrical Infrastructure), and Chair of the EPFL Energy Centre Directorate.
	He has authored or co-authored over 260 scientific papers published in reviewed journals and international conferences.
	His research interests include power systems with particular reference to real-time monitoring and operation, power system protections, power systems dynamics, and power system transients.
	Dr. Paolone was the Co-Chairperson of the TPC of the 2009 edition of the International Conference of Power Systems Transients, Vice-Chair and Chair of the TPCs of the 2016 and 2018 editions of the Power Systems Computation Conference.
	In 2013, he was the recipient of the IEEE EMC Society Technical Achievement Award.
	He has co-authored several papers that received the awards in mainstream power systems journals and conferences.
	He is the Editor-in-Chief of the journal Sustainable Energy, Grids and Networks (Elsevier) and Associate Editor of the IEEE Transactions on Industrial Informatics. 
\end{IEEEbiography}